%% file: try.tex
\documentclass[aps,prl,twocolumn]{revtex4}
\usepackage{epsfig,amssymb,amsmath,amsthm,amsfonts,amsbsy,mathrsfs,pifont}
\usepackage{graphicx}
\usepackage{color}
\usepackage{tikz}
\usetikzlibrary{arrows, shapes, chains}
\usetikzlibrary{plotmarks}
\newcommand{\hollowstar}{\text{\ding{73}}}
\newcommand\marksymbol[2]{\tikz[#2,scale=1.2]\pgfuseplotmark{#1};}
\usepackage{xcolor,hyperref}
\usepackage[export]{adjustbox}

\begin{document}
\title{Two-Dimensional Melting of two- and three-component mixtures}
\author{Yan-Wei Li}
\email{yanweili@bit.edu.cn}
\affiliation{Key Laboratory of Advanced Optoelectronic Quantum Architecture and Measurement (MOE), School of Physics, Beijing Institute of Technology, Beijing, 100081, China}
\author{Yugui Yao}
\email{ygyao@bit.edu.cn}
\affiliation{Key Laboratory of Advanced Optoelectronic Quantum Architecture and Measurement (MOE), School of Physics, Beijing Institute of Technology, Beijing, 100081, China}
\author{Massimo Pica Ciamarra}
\email{massimo@ntu.edu.sg}
\affiliation{Division of Physics and Applied Physics, School of Physical and
Mathematical Sciences, Nanyang Technological University, Singapore 637371, Singapore}
\affiliation{
CNR--SPIN, Dipartimento di Scienze Fisiche,
Universit\`a di Napoli Federico II, I-80126, Napoli, Italy
}
\affiliation{CNRS@CREATE LTD, 1 Create Way, \#08-01 CREATE Tower, Singapore 138602}
\date{\today}

\begin{abstract}
We elucidate the interplay between diverse two-dimensional melting pathways and establish solid/hexatic and hexatic/liquid transition criteria via the numerical simulations of the melting transition of two- and three-component mixtures of hard polygons and disks.
We show that a mixture's melting pathway may differ from its components and demonstrate eutectic mixtures that crystallize at a higher density than their pure components.
Comparing the melting scenario of many two- and three-component mixtures, we establish universal melting criteria: the solid and hexatic phases become unstable as the density of topological defects respectively overcomes $\rho_{d,{\rm s}} \simeq 0.046$ and $\rho_{d,{\rm h}}\simeq 0.123$.
\end{abstract}
\maketitle

The nature of the melting of two-dimensional (2d) solids is a long-standing and fascinating problem in statistical physics~\cite{KT, HN, Y,Brinkman1982, Heiney, Strandburg,LJpd1981,Krauth2011, Experiment_harddisc, Massimo_2020PRLmelting, Tanaka,Cell, pinned_particle,Smith2021,Keim,Daniel2022}.
According to the Kosterlitz-Thouless-Halperin-Nelson-Young (KTHNY) theory~\cite{KT, HN, Y}, 2d solids melt via a continuous solid to hexatic transition driven by the unbinding of dislocation pairs, followed by a continuous hexatic to liquid transition driven by the further unbinding of isolated dislocation into disclinations.
Two alternative melting scenarios are possible.
In the mixed case, the solid-to-hexatic transition is continuous, while the hexatic-to-liquid transition is discontinuous;
Finally, the solid transforms into a liquid via a first-order transition in the discontinuous case.
Recent computational and methodological advances have allowed establishing the melting scenario of pure two-dimensional substances~\cite{Krauth2011,
Krauth2015, Glotzer, pinned_particle, Santi, ningxu, Tanaka,Cell, pinned_particle, Smith2021, Keim, Daniel2022, Experiment_harddisc, Massimo_2020PRLmelting}.
It has been clarified, for instance, that shape controls the melting scenario in hard-particle systems~\cite{Krauth2011, Experiment_harddisc, Massimo_2020PRLmelting} with, e.g., hexagons, disks and pentagons following the KTHNY, mixed and discontinuous melting~\cite{Glotzer, Massimo_2020PRLmelting}.
In soft repulsive systems, the softness of the inter-particle interaction influences the melting pathway~\cite{Krauth2015}, while attractive interactions promote discontinuous melting~\cite{Massimo_2020PRLmelting}.

Phase diagrams are more complex in many-component systems.
The relative fractions become an essential control variable, novel phases may emerge, and melting may interplay with phase separation.
These aspects have been thoroughly investigated in three-dimensional systems but are
almost unexplored in two dimensions~\cite{John_Russo, Guo2021, Ran2dmelt}.
Can a mixture melt via a scenario different from its components?
Is there a eutectic mixture whose melting temperature is lower than its components or, for hard particles, a mixture that melts at a density higher than its components?
Do topological defects correlate with the phase behaviour as in pure systems~\cite{defects_sm}?
Investigating the melting of mixtures further offers the possibility of validating proposed melting criteria based on a generalization of Lindemann's approach~\cite{Zahn2000, Khrapak2020, Dillmann2012} or topological defects~\cite{defects_sm, Guo2021}.

In this Letter, we investigate the interplay between diverse melting mechanisms by tuning the relative composition of two- and three-component mixtures of hard particles.
In hexagons+disks and pentagons+hexagons mixtures, the melting pathway smoothly crossovers from that of a pure substance to that of the other.
On the contrary, disks+pentagons mixtures may follow the KTHNY melting scenario, not occurring in pure disks or pentagons.
The study of a large catalogue of two-dimensional systems further demonstrates that the density of topological defects $\rho_d$ controls the stability of the solid and hexatic phases.
As the density decreases, the solid phase becomes unstable at $\rho_{d,{\rm s}} \simeq 0.046$, and the hexatic one at $\rho_{d,{\rm h}}\simeq 0.123$~\cite{Guo2021}.
Our predictions open new routes to self-assembly in two dimensions~\cite{Zong2022} and can be experimentally verified by changing a mixture's composition.

\begin{figure*}[!t]
 \centering
 \includegraphics[angle=0,width=1\textwidth]{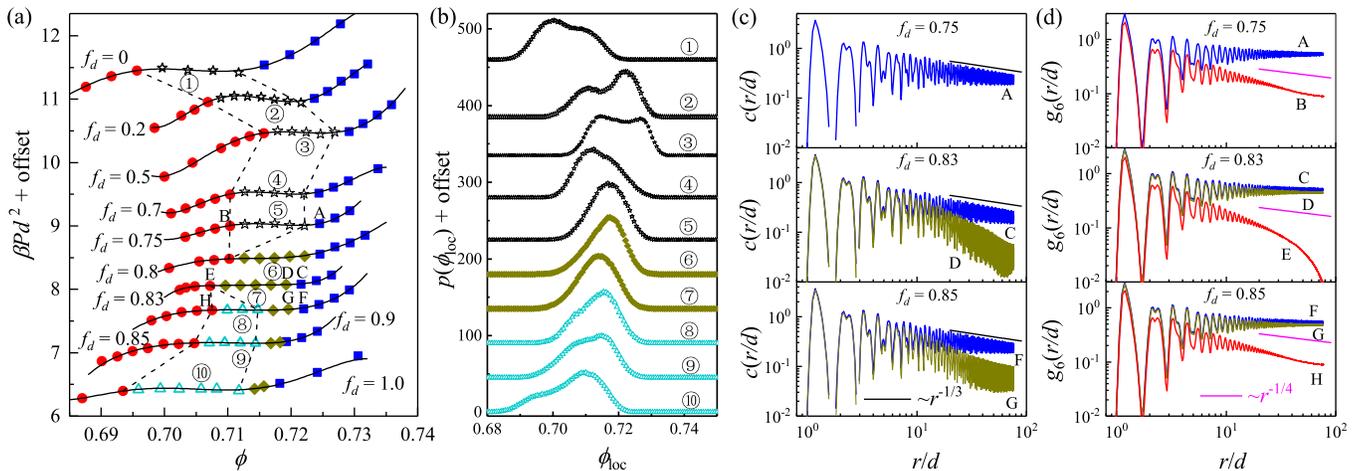}
 \caption{
Melting of disks+pentagons mixtures.
(a) Isothermal equation of state. Data are vertically shifted for clarity, and the fraction of disks increases from top to bottom.
(b) probability distribution of the local density, and (c) the translational and (d) the bond-orientational correlation functions for selected state points marked in (a).
Symbols and colours identify different phases: liquid (\protect\marksymbol{*}{red}); hexatic (\protect\marksymbol{diamond*}{olive}); solid (\protect\marksymbol{square*}{blue}); liquid/solid coexistence (\hollowstar); liquid/hexatic coexistence (\protect\marksymbol{triangle}{cyan}).
\label{fig:mixpd}
}
\end{figure*}

We construct pentagons, hexagons and disks with the same circumscribing circle of diameter $d$, by lumping together $N_d=40$ or $42$ beads equally spaced along the perimeter, as detailed in the Supplemental Material (SM)~\cite{SM} and in ~\cite{Massimo_2020PRLmelting}.
Beads of different particles with separation distance $r_{\rm b}$ interact via the Weeks, Chandler, and Andersen (WCA) potential~\cite{wca}: $V_{\rm WCA}(r_{\rm b}) = 4\epsilon \left[ \left(\frac{\sigma}{r_{\rm b}}\right)^{12}- \left(\frac{\sigma}{r_{\rm b}}\right)^{6} + c\right]$ for $r_{\rm b} \leq r_{\rm cut} = 2^{1/6}\sigma$. 
The constant $c$ enforces $V_{\rm WCA}(r_{\rm cut}) = 0$. 
We set $\sigma\simeq 0.14d$. 
We carry out molecular dynamics simulations in the canonical ensemble, under periodic boundary conditions within a rectangular box with a side length ratio of $2:\sqrt{3}$ to accommodate the triangular lattice, fixing the temperature via the Nos$\rm {\acute{e}}$-Hoover thermostat.
We fix $T=20\epsilon/k_{\rm b}$ if not otherwise specified and demonstrate in Fig. S4~\cite{SM} that the choice of the temperature values has no effect on the melting scenario.
All simulations are performed using the graphics processing unit-accelerated \texttt{GALAMOST} package~\cite{Galamost}. 
We ensure thermal equilibration as detailed in Fig. S2~\cite{SM}, and report results for systems of $N=20521$ particles, where finite size effects are negligible as detailed in Fig. S3~\cite{SM}.

Most previous investigations of the equilibrium phase diagram of two-dimensional mixtures of particles interacting via purely repulsive forces considered the specific case of hard disks.
These studies focused on identifying the many possible crystalline phases determined by the size ratio and the relative fraction, e.g.~\cite{Likos93, UCHE2004428, Castaeda2016, Smallenburg2020}, and on the stability of the hard-disk melting scenario~\cite{John_Russo, Guo2021, Ran2dmelt}.
In these systems, geometric frustration stemming from the size disparity prevents the triangular lattice from being the ground state.
We depart from these previous works as we consider mixtures of particles with similar sizes but different shapes. By focusing on mixtures of particles with similar sizes, all crystallizing in the triangular lattice with the same lattice constant, we avoid phase separation~\cite{Escobedo2021}, glass formation~\cite{Speedy1999} and the emergence of complex crystalline phases on increasing the density.
Yet, the shapes we consider melt via different pathways.
Hence, while previous works considered the competition between different possible ground states, we explore the competition between alternative crystallization pathways to the same ground state.

To determine the melting scenario, we first consider if the isothermal equation of state (EOS) possesses a Mayer-Wood loop indicating the presence of a first-order transition~\cite{Mayer_wood}.
If so, we fit the EOS to a fifth-order polynomial to identify the coexistence boundaries and further investigate the local density distribution in the coexisting region~\footnote{We associate to each particle a local density defined as $\phi_{\rm loc}(\vec{r}_i)=\frac{\sum_{j=1}^{N}A_jH(r_c-|\vec{r}_{i}-\vec{r}_{j}|)}{\pi r_c^2}$, where $A_j$ is the area of particle $j$, $H$ is the Heaviside step function, and $r_c=21d$.}.
Outside the coexistence region, or in its absence, we determine the pure phases by investigating the correlation functions $c(r)$ and $g_6(r)$ of the translational and bond-orientational order~\cite{KT, Krauth2011, Massimo_PRE2019}.
In the liquid phase, both functions decay exponentially.
In the hexatic phase~\cite{KT}, the system possesses  quasi-long-range bond-orientational order and short-range translational order, so that $g_{6}(r)\propto r^{-\eta_{6}}$ with $\eta_{6}<1/4$, while $c(r)$ decays exponentially.
In the solid phase, $g_{6}(r)$ does not decay as the bond-orientational order is extended, while $c(r)\propto r^{-\eta}$ with $\eta \le 1/3$ decay algebraically as Mermin-Wagner fluctuations makes the translational order quasi-long-ranged~\cite{Mermin}.

\begin{figure*}[!t]
 \centering
 \includegraphics[angle=0,width=1\textwidth]{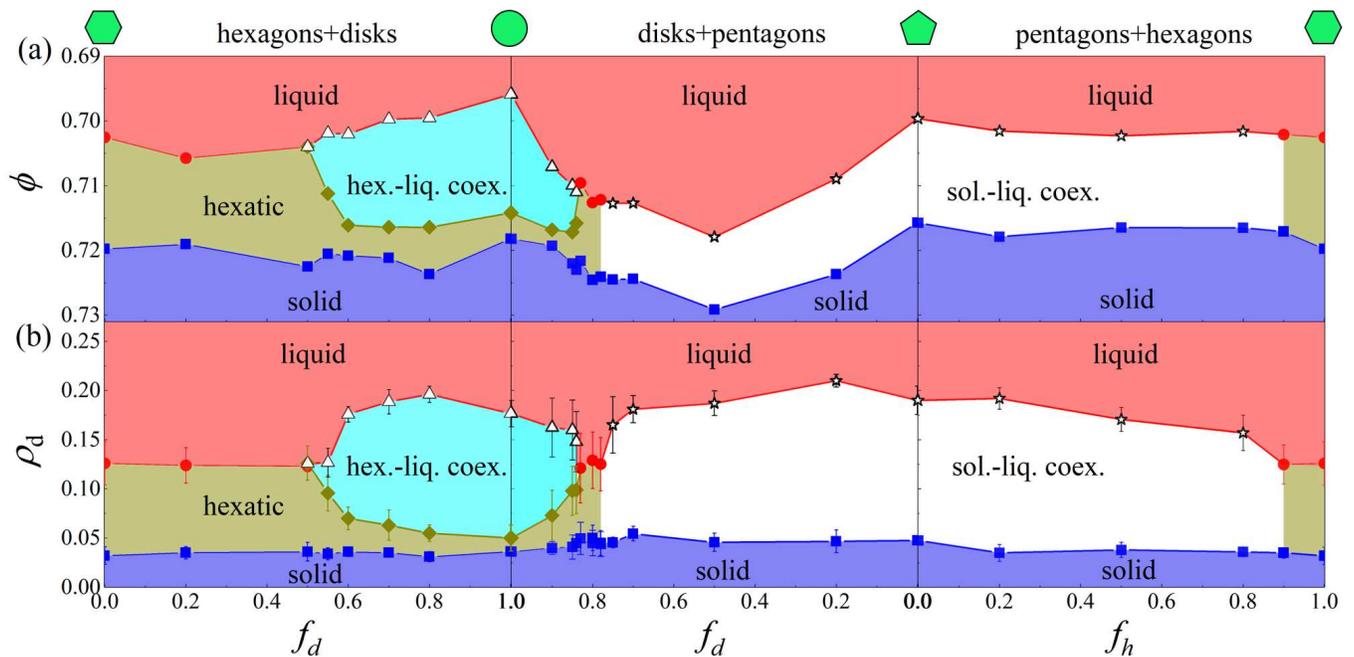}
  \caption{
(a): Phase diagram of hexagons+disks, disks+pentagons, and pentagons+hexagons mixtures, in the relative fraction $f$ and area fraction $\phi$ plane.
(b): the same diagrams are illustrated in the $f$-$\rho_d$ plane, with $\rho_d$ the density of defective particles that do not have six immediate neighbours. We evaluate $\rho_d$ in $8$ sub-blocks of the system to estimate its error.
The liquid/hexatic, solid/hexatic, and solid/coexistence boundaries occur at approximately constant defect densities.
\label{fig:pd}
}
\end{figure*}

Figure~\ref{fig:mixpd} illustrates the dependence of the melting scenario of disks+pentagons mixtures on the fraction of disks, $f_d$.
The equation of state (a) reveals a first-order transition at both small and large $f_d$ that corresponds to the discontinuous solid-liquid transition of pure pentagons and the discontinuous liquid-hexatic transitions of pure disks.
When the transition is discontinuous, the local density distribution for state points inside the coexistence region (panel b) shows clear bimodality. Consistently with the absence of segregation phenomena, the coexisting phases have the same disk fraction $f_d$.

The translational and bond-orientational correlation functions illustrated in Figs.~\ref{fig:mixpd}(c) and \ref{fig:mixpd}(d) clarify that on increasing the disk fraction, the discontinuous hard-pentagons melting scenario first transforms into the continuous KTHNY scenario and then into that of hard-disk.
Indeed, we find melting is continuous for disk fractions around $f_d \simeq 0.82$.
This result demonstrates that a two-component mixture may melt via a scenario different from its components.
We notice that the crossover from a mixed to discontinuous melting may not involve an intermediate KTHNY melting scenario, as observed in attractive hard disks~\cite{Massimo_2020PRLmelting} as the temperature decreases.

We have also investigated the melting scenario of hexagons+disks and pentagons+hexagons mixtures and provide details in the SM~\cite{SM}.
We summarize our investigations in Fig.~\ref{fig:pd} by illustrating how the melting scenario of hexagons+disks, disks+pentagons and pentagons+hexagons varies with the fraction of disks, $f_d$, and that of hexagons, $f_h$.
The three diagrams match at their boundaries corresponding to the melting behaviour of pure disks, pure pentagons and pure hexagons, which we find to follow the mixed, discontinuous and KTHNY scenario, consistently with previous results~\cite{Krauth2011, Glotzer, Massimo_2020PRLmelting}.

Hexagons+disks mixtures (panel a, left) crossover from the KTHNY to the mixed scenario as the fraction of disks overcomes $f_d \simeq 0.5$, an intermediate value suggesting that these two shapes do not frustrate each other considerably.
Conversely, pentagons+hexagons mixtures (panel a, right) crossover from the discontinuous to the KTHNY scenario as the fraction of hexagons overcomes $f_h \simeq 0.9$: adding a small fraction of pentagons disrupts hexagons' KTHNY melting scenario.
Similarly, a small amount of pentagons disrupts the melting scenario of hard disks (panel a, middle).
However, in this case, on increasing the pentagons fraction, the melting scenario first becomes of KTHNY type and then becomes discontinuous, as we previously noticed.

In hexagons+disks and pentagons+hexagons mixtures, the stability limit $\phi_L(f)$ of the liquid phase weakly depends on $f$.
Conversely, in disks+pentagons mixtures, for $f \simeq 0.5$, the liquid phase is stable up to $\phi_L \simeq 0.715$, a value sensibly higher than that characterizing the pure phases, $\phi_L \simeq 0.70$.
This mixture is thus one of the few known examples of eutectic two-dimensional colloidal systems~\cite{Toyotama2016}.

Melting in two dimensions has historically been related to the evolution of the density of topological defects, clusters of particles not having $6$ neighbours.
In the KTHNY scenario~\cite{KT, HN, Y}, melting occurs through a continuous solid-hexatic transition driven by the unbinding of dislocation pairs (clusters of 4 particles with 5+7+5+7 neighbours), followed by a continuous hexatic-liquid one driven by the further unbinding of isolated dislocation (5+7) into disclinations (5 or 7).
While the observation of extended clusters of defective particles~\cite{defects_sm} suggests a rather complex relation between phase behaviour and defects, Guo et al.~\cite{Guo2021} have found that in systems following the KTHNY scenario, at the hexatic/liquid boundary, the density of defective particles acquires a universal value, $\rho_{d,{\rm h}}= 0.12$.
This number then acts as an upper bound for the defects' density in the hexatic phase.

We explore the relationship between defects' density and phase behaviour by investigating the phase diagrams in the $\rho_d-f$ plane in Fig.~\ref{fig:pd}(b).
We find that the defects' density at the hexatic/liquid boundary (red circles) attains the universal value of $\rho_{d,{\rm h}} = 0.123\pm 0.006$, in agreement with the earlier speculation~\cite{Guo2021}.
In addition, we also find the defects' density equals $\rho_{d,{\rm s}} = 0.046\pm0.005$ at the solid/hexatic and solid/coexistence boundaries.

\begin{figure}[!t]
 \centering
 \includegraphics[angle=0,width=1.0\columnwidth]{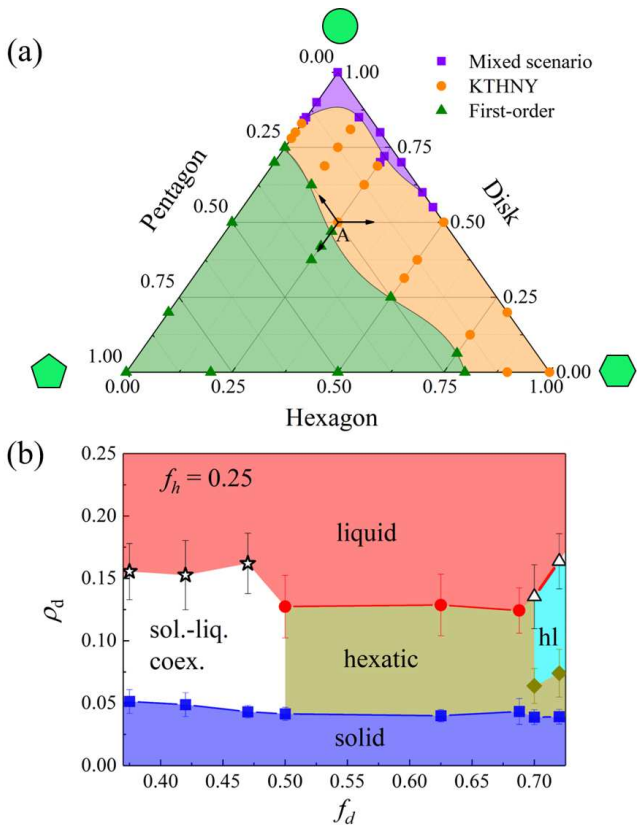}
\caption{
(a)
Ternary phase diagram for mixtures of pentagons, hexagons and disks.
Each point within the triangle corresponds to a composition determined by the intersections of lines passing through that point and the sides of the triangle. For example, in point $A$, the fraction of disks, pentagons,  and hexagons are $0.50$, $0.25$ and $0.25$.
(b)
Phase diagram at fixed hexagon fraction $f_h=0.25$, in the disk fraction, defect density plane.
$\rho_d$ acquires constant values at the solid/hexatic and hexatic/liquid boundaries.
\label{fig:ternary}}
\end{figure}
To test the universality of these findings, we examine ternary mixtures of pentagons, hexagons and disks.
For each composition, we investigate the phase behaviour as a function of the volume fraction to determine the melting scenario~\cite{SM}.
The ternary phase diagram of Fig.~\ref{fig:ternary}(a) summarizes the result of this investigation.
Each point in the diagram represents the melting scenario observed at a given composition $f_d,f_p,f_h$.
A point's composition is read by connecting it to the triangle's sides, as illustrated with point `A' for which $f_d=0.50$, $f_p=0.25$ and $f_h=0.25$.
Hence, the three corners correspond to the three pure systems and the three sides to the three binary diagrams, e.g., the bottom left corner corresponds to pure pentagons and the `pentagon' side to the disks+pentagons mixture.

Pentagon's discontinuous melting scenario extends over a large part of the diagram: melting becomes discontinuous as soon as the pentagon's fraction overcomes $f_h \simeq 0.25$.
The KTHNY scenario always separates the mixed and discontinuous transitions regardless of the hexagon fraction.

In bidisperse systems, the defects' density controls the stability of the solid and hexatic phases.
We have found the same occurs in the ternary mixtures.
To demonstrate this, we focus here on ternary mixtures with a fixed fraction of hexagons, $f_h = 0.25$, and report analogues results for other $f_h$ values in the SM~\cite{SM}.
For the studied fraction of disks, $f_d \leq 1-f_h$, we plot in Fig.~\ref{fig:ternary}(b) the value of $\rho_d$ at the boundaries between the different phases the system traverses as the area fraction increases.
Regardless of the $f_d$ values, the solid and hexatic phases become unstable at $\rho_{d,{\rm s}}\simeq 0.046$ and $\rho_{d,{\rm h}}\simeq 0.123$, respectively, as observed in the binary mixtures and the pure phases.

We have studied the two-dimensional melting of mixtures of similar-sized hard disks, hexagons and pentagons.
The size similarity suppresses geometric frustration and ensures crystallization in the triangular lattice for all compositions.
As such, changes in the relative composition allow us to investigate the interplay between the discontinuous melting scenario of hard pentagons, the mixed scenario of hard disks, and the KTHNY scenario of hard hexagons.
Hexagons+disks mixtures smoothly crossover from the KTHNY to the mixed melting scenario; similarly, hexagons+pentagons mixtures crossover from the KTHNY to the discontinuous melting scenario.
On the contrary, in disks+pentagons mixtures, the KTHNY melting may result from the competition between hard disks' mixed melting scenario and pentagons' discontinuous one: two dogs strive for a bone, and the third runs away with it.
The exhaustive investigation of the melting of two- and three-component mixtures further corroborates speculated~\cite{Guo2021} melting criteria and establishes new ones based on the density of defects.
The solid phase becomes unstable for $\rho_d > \rho_{d,{\rm s}}$, the hexatic phase for $\rho_d > \rho_{d,{\rm h}}$.
The connection of these criteria with previously speculated Lindemann-like ones~\cite{Zahn2000, Khrapak2020, Dillmann2012} and the possibility of relating the free energy of the phases to the defect's density are fascinating future research directions that could lead to a better understanding of the mechanisms driving the different melting pathways.

Y.-W.L. acknowledges support of the start-up funding of Beijing Institute of Technology and support
from the National Natural Science Foundation (NSF) of China (Grants No. 12105012). Y.Y. is supported by the NSF of China (Grants Nos. 11734003, 12061131002).
M.P.C.  acknowledges support by the Ministry of Education, Singapore, under its
Academic Research Fund Tier 2 (MOE-T2EP50221-0016) and Tier 1 (MOE-T1-RG56/21).
\bibliographystyle{apsrev4-1}
\input{try.bbl}
\end{document}


\title{Supplemental Material \\ for \\
  Two-Dimensional Melting of two- and three-component mixtures
}

\author{Yan-Wei Li}
\affiliation{Key Laboratory of Advanced Optoelectronic Quantum Architecture and Measurement (MOE), School of Physics, Beijing Institute of Technology, Beijing, 100081, China}
\author{Yugui Yao}
\affiliation{Key Laboratory of Advanced Optoelectronic Quantum Architecture and Measurement (MOE), School of Physics, Beijing Institute of Technology, Beijing, 100081, China}
\author{Massimo Pica Ciamarra}
\affiliation{Division of Physics and Applied Physics, School of Physical and
Mathematical Sciences, Nanyang Technological University, Singapore 637371, Singapore}
\affiliation{
CNR--SPIN, Dipartimento di Scienze Fisiche,
Universit\`a di Napoli Federico II, I-80126, Napoli, Italy
}
\maketitle

\onecolumngrid
\tableofcontents
~\newpage
\section{Numerical details \label{sec:detail}}
\subsection{Model}

\begin{figure*}[!b]
 \centering
 \includegraphics[angle=0,width=0.95\textwidth]{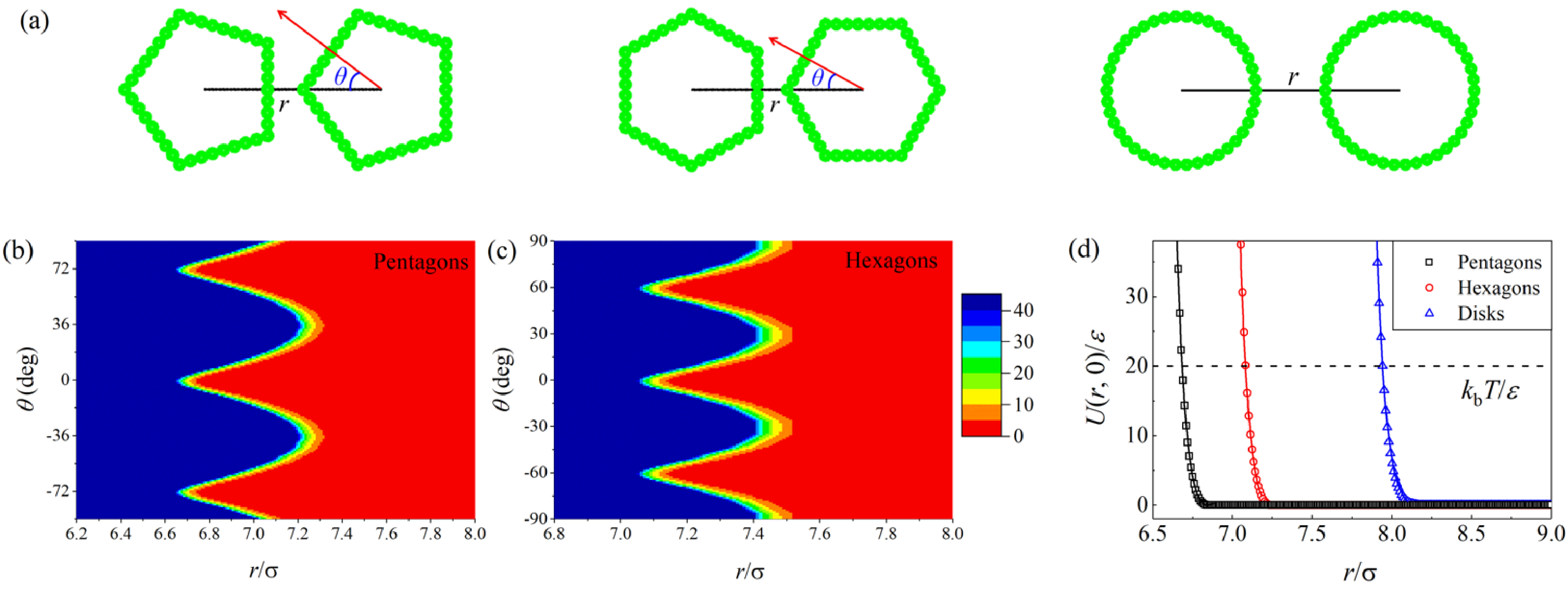}
 \caption{
(a) Illustration of the numerical model. 
Pentagons, hexagons and disks are rigid bodies constructed by aggregating point particles (beads) along their perimeter. Beads of different bodies interact via the WCA potential.
Panel (b) [(c)] illustrates the interaction energy $U(r,\theta)$ between pentagons [hexagons], as a function of the interparticle separation $r$ and relative orientation $\theta$.
Panel (d) shows the interaction energy $U(r,\theta=0)$ between bodies with relative orientation $\theta=0$. 
The dashed line marks the temperature investigated in the main text.
The role of temperature is addressed in Sec.~\ref{sec:tem}.
\label{fig:model}
}
\end{figure*}

We model extended pentagons, hexagons, and disks as a collection of $N_d$ rigidly connected point particles (beads) uniformly spaced along the shapes' perimeters, as shown in Fig.~\ref{fig:model}(a). 
Beads of different extended particles at a separation distance $r_{\rm b}$ interact via the purely repulsive Weeks, Chandler, Andersen (WCA) potential~\cite{wca}: $V_{\rm WCA}(r_{\rm b}) = 4\epsilon \left[ \left(\frac{\sigma}{r_{\rm b}}\right)^{12}- \left(\frac{\sigma}{r_{\rm b}}\right)^{6} + c\right]$ for $r_{\rm b} \leq r_{\rm cut} = 2^{1/6}\sigma$,  $V_{\rm WCA}(r_{\rm b}) = 0$ otherwise. 
The constant $c$ ensures that $V_{\rm WCA}(r_{\rm cut}) = 0$.

The shapes have the same circumcircle diameter $d=7.06\sigma$.
This choice ensures particles are similarly sized and prevents phase separation. 
Indeed, we have verified that in A+B mixtures, the distribution of the fraction of neighbours of type $A$ of any given particle is density-independent and peaks at $f_A/(f_A+f_B)$. 

The interaction energy $U(r,\theta)$ between two extended particles is the sum of interactions between their constituent beads, and depends on the separation distance $r$ and relative orientation angle $\theta$ defined in Fig.~\ref{fig:model}(a).
The angular dependence of the interaction energy between $n$-sided particles exhibits rotational symmetry, with $U(r,\theta)=U(r,\theta+\frac{2\pi}{n})$. 
Figs.~\ref{fig:model}(b) and \ref{fig:model}(c) illustrate $U(r,\theta)$ for pentagons and hexagons.

When two $n$-sided polygonal particles are at a separation $r$ with $\theta = 0$, the distance between the beads of opposing faces is $r-d_{\rm in}$, with $d_{\rm in} = d\cos(\pi/n)$ the diameter of the inscribed circle. 
Since each edge of an $n$-sided polygon contains $n_b = 2+(N_d-n)/n$ beads, close to contact the interaction energy is $U(r,\theta= 0) = n_b V_{\rm WCA}(r-d_{\rm in})$. 
For disks, $n=N_d$.
We illustrate $U(r,\theta= 0)$ in Fig.~\ref{fig:model}(d) and summarize the values of the relevant parameters in Table~\ref{tb:parameters}.

\begin{table}[!!!b]
    \centering
\begin{tabular}{|l |c |c | c |}
 \hline &~Pentagons~&~Hexagons~&~Disks~\\ 
 \hline $n_b$ & 9 & 8 & 2 \\
 \hline $d_{\rm in}/\sigma$ & 5.71 & 6.11 & 7.04   \\
 \hline $N_d$ & 40 & 42 & 42   \\
 \hline $r_T(T=10\epsilon/k_{\rm b})$ & 6.72 & 7.11 & 7.98   \\
 \hline $r_T(T=20\epsilon/k_{\rm b})$ & 6.68 & 7.08 & 7.94   \\
 \hline $r_T(T=50\epsilon/k_{\rm b})$ & 6.63 & 7.03 & 7.89   \\
 \hline
\end{tabular}
\caption{Number of beads on each edge, $n_b$, the diameter of the inscribed circle, $d_{\rm in}$, total number of beads $N_d$, and thermal diameter $r_T$ of the inscribed circle, for the different extended particles. 
\label{tb:parameters}
}
 \end{table}

\subsection{Units}
We use the diameter $d$ of the circumscribed circle as a unit of length, $\epsilon$ as the unit of energy, and the mass $m$ of the beads used to construct the extended particles as the unit of mass. 

To facilitate comparison with hard-sphere colloidal experiments, we express the density of the system in terms of the area fraction $\phi$.
To evaluate the area fraction, we consider that at temperature $T$ the effective diameter $r_T(T)$ of the circle inscribed within each shape satisfies $U(r_T,\theta=0)=k_{\rm b}T$. 
Henceforth, we assume hexagonal and pentagonal particles have the area of hexagons and pentagons with an inscribed circle of diameter $r_T$, and disks have area $\pi (r_T/2)^2$.
Table~\ref{tb:parameters} reports the $r_T$ values at the considered temperatures.

\subsection{Method}
We conduct molecular dynamics simulations in the NVT ensemble using periodic boundary conditions.
The number of extended particles varies from $N=1243$ to $N=51643$, and $N=20521$ unless otherwise specified. 
We integrate the equations of motion using a Verlet algorithm~\cite{Allen_book} and maintain a fixed temperature of $T=20\epsilon/k_{\rm b}$, if not otherwise specified, via a Nos$\rm {\acute{e}}$-Hoover thermostat. All simulations are performed using the GPU-accelerated GALAMOST package~\cite{Galamost}.

\section{Thermal equilibration \label{sec:Checkeq}}
\begin{figure*}[!h]
 \centering
 \includegraphics[angle=0,width=0.88\textwidth]{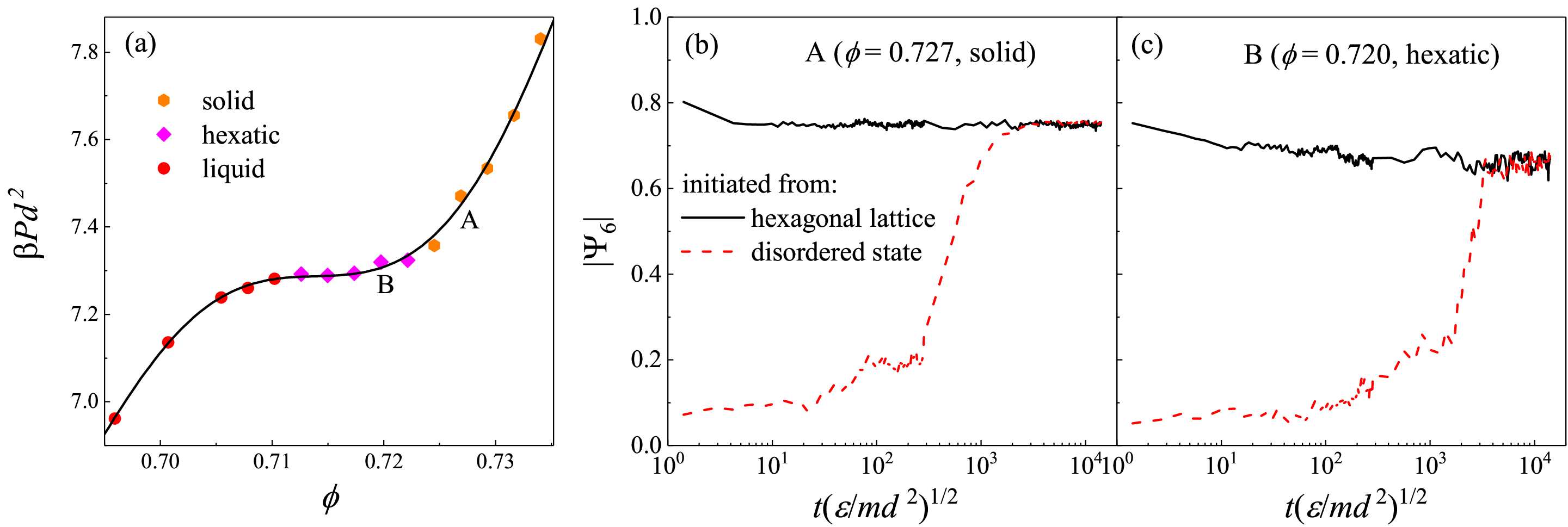}
 \caption{
(a) Equation of state of the pentagons+disks mixture at $f_d=0.8$.
(b,c) Time evolution of the global order parameter for the state points in the solid and hexatic phases marked with A and B in (a).
\label{fig:eq}
}
\end{figure*}

The data reported in this work are obtained by preparing the systems in a mixed crystalline state at a high area fraction.
We then decrease the area fraction in small steps of $\Delta\phi \simeq 0.0005$, and at each investigated area fraction, we collect data after simulations lasting at least $t_{\rm eq}=5\times10^3 \sqrt{md^2/\epsilon}$.
To prove $t_{\rm eq}$ allows the system to equilibrate, we consider the evolution of the order parameter $\Psi_6$ for a pentagons+disks mixture at state point A ($\phi = 0.727$) and B ($\phi = 0.720$).
In equilibrium, our study suggests these state points are in the hexatic and solid phases as illustrated in the equation of state in Fig.~\ref{fig:eq}(a). 
We compare simulations initiated from a hexagonal lattice and a disordered configuration.
Figs.~\ref{fig:eq}(b) and \ref{fig:eq}(c) show that the order parameter reaches the same asymptotic value regardless of the initial condition in a time smaller than $t_{\rm eq}$, proving our simulations are long enough to reach thermal equilibrium.
We also verified that, after our considered equilibration period, the dynamics does not depend on the system's initial configuration.
These data are illustrative of analogous results we find for other densities and mixtures.

\newpage
\section{Finite size effects \label{sec:size}}
We investigate the size dependence of the different melting scenarios focusing on the pentagons+disks mixture at $f_d=0.2$, $0.83$, and $0.9$. 
At these $f_d$ values, for $N=20521$ melting appears to follow the first-order, the KTHNY, and the mixed scenario, respectively.

\begin{figure*}[!!!h]
 \centering
 \includegraphics[angle=0,width=0.95\textwidth]{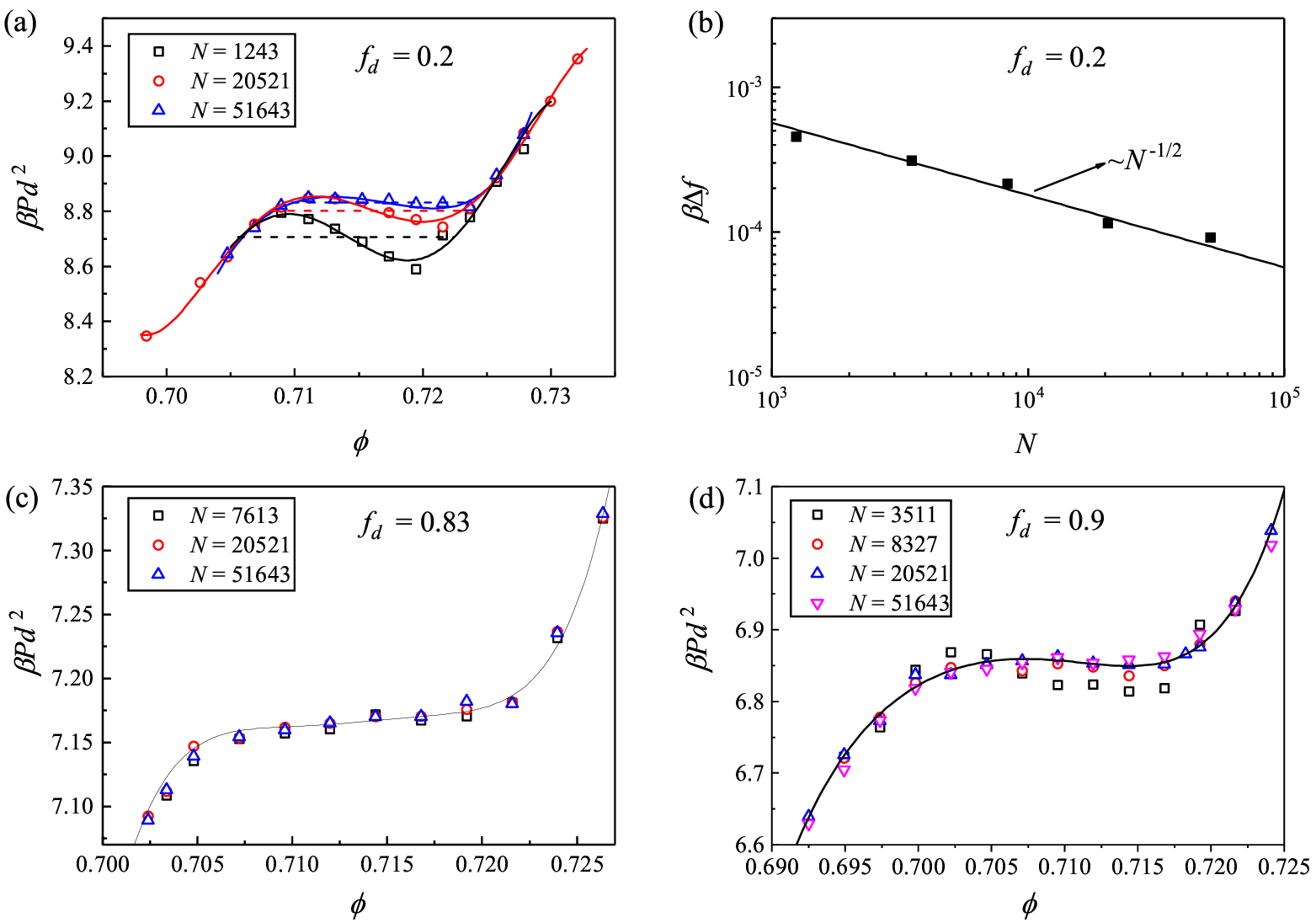}
 \caption{
 Equation of state of pentagons+disks mixtures at $f_d=0.2$ (a), $0.83$ (c) and $0.9$ (d), for different values of the system size $N$.
 The black solid lines correspond to fifth-order polynomial fits. Panel (b) illustrates the size dependence of the interfacial free energy for $f_d=0.2$.
\label{fig:size}
}
\end{figure*}

At $f_d=0.2$ (Fig.~\ref{fig:size}(a)), the amplitude of the Mayer-Wood loops scales as $N^{-1/2}$ (Fig.~\ref{fig:size}(b)), as expected for a discontinuous transition. 
At $f_d=0.83$, the equation of state has no Mayer-Wood loop and system size dependence, as we illustrate in Fig.~\ref{fig:size}(c), confirming the system melts via the continuous KTHNY transitions. 
At $f_d=0.9$, the size dependence disappears for moderate system size $N\ge8327$.
These results indicate that our $N=20521$ results are representative of the thermodynamic limit.

\newpage
\section{Temperature dependence of the melting scenario\label{sec:tem}}
\begin{figure*}[!h]
\centering
\includegraphics[angle=0,width=0.95\textwidth]{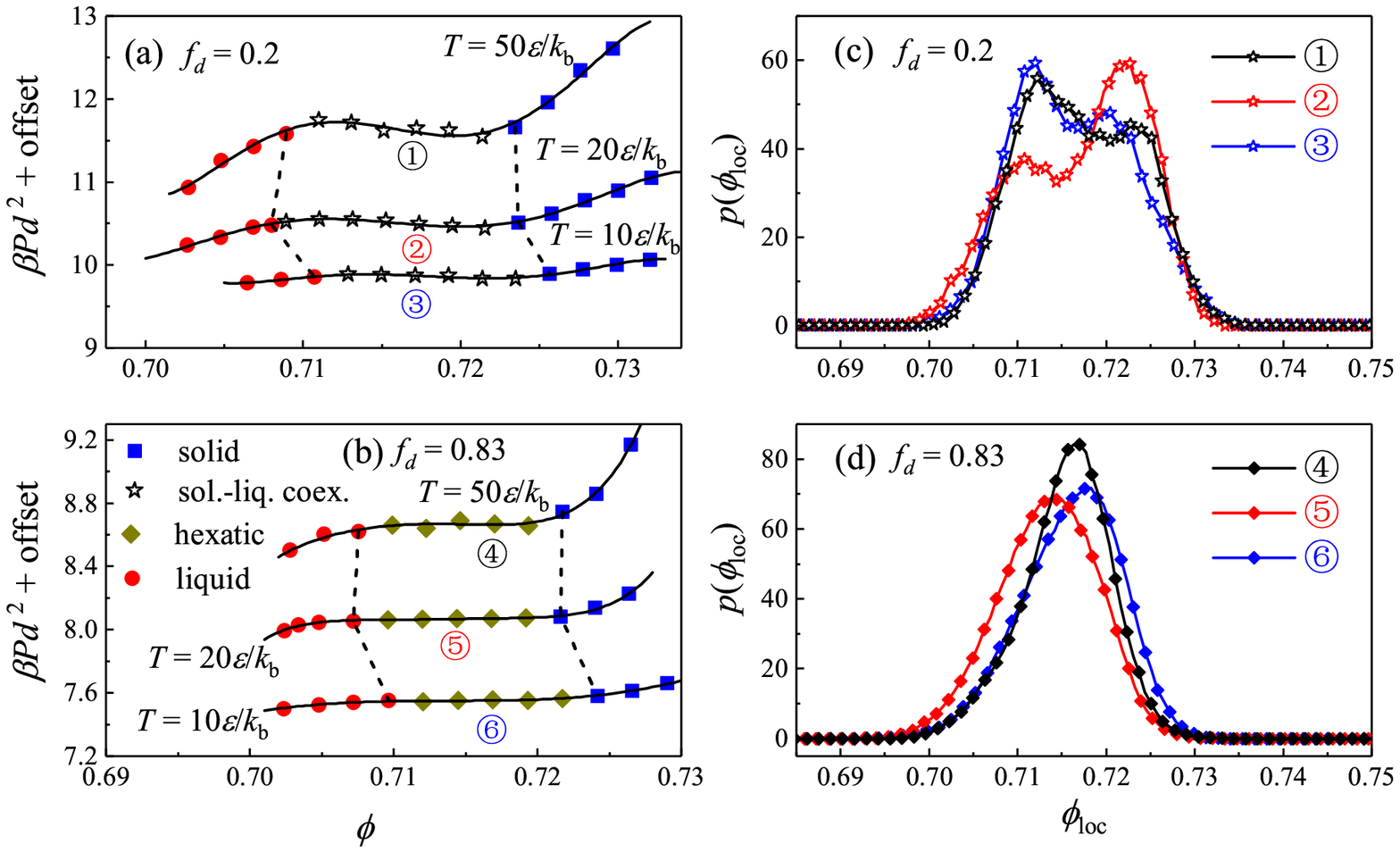}
\caption{Temperature dependence of the equation of state for pentagons+disks mixtures at (a) $f_d=0.2$ and (b) $f_d=0.83$. 
Panels (c) and (d) illustrate the probability distribution of the local density at the state points marked in (a) and (b), respectively.
\label{fig:Tdepend}
}
\end{figure*}

We report in the main manuscript results obtained at $T=20\epsilon$. 
Here we investigate how temperature influences the melting scenario by focusing on pentagons+disks mixtures with $f_d=0.2$ and $f_d=0.83$.
We compare simulations at $T=10\epsilon/k_{\rm b}$, $T=20\epsilon/k_{\rm b}$, and $T=50\epsilon/k_{\rm b}$.
At $f_d=0.2$, we observe a Mayer-Wood loop (Fig.~\ref{fig:Tdepend}(a)) and a bimodal distribution of the local density (Fig.~\ref{fig:Tdepend}(c)) indicating the melting transition is first-order. 
The distribution slowly evolves as coarsening proceeds, with complete phase separation occurring on a time scale not accessible to our simulation.
For $f_d=0.83$, we find a KTHNY melting scenario, supported by the absence of a pressure loop and a unimodal distribution of the local density (Figs.~\ref{fig:Tdepend}(b) and \ref{fig:Tdepend}(d)).
The phase boundaries are weakly temperature dependent and almost coincide at the highest temperatures.
These results have a weak temperature dependence, suggesting they represent the hard particle limit.

\newpage
\section{Melting of bidisperse mixtures \label{sec:binary}}
In the main text, we identified the melting scenario of pentagons+disks mixtures.
Here we describe the melting of disks+hexagons and hexagons+pentagons mixtures. 

\begin{figure}[!h]
 \centering
 \includegraphics[angle=0,width=0.75\textwidth]{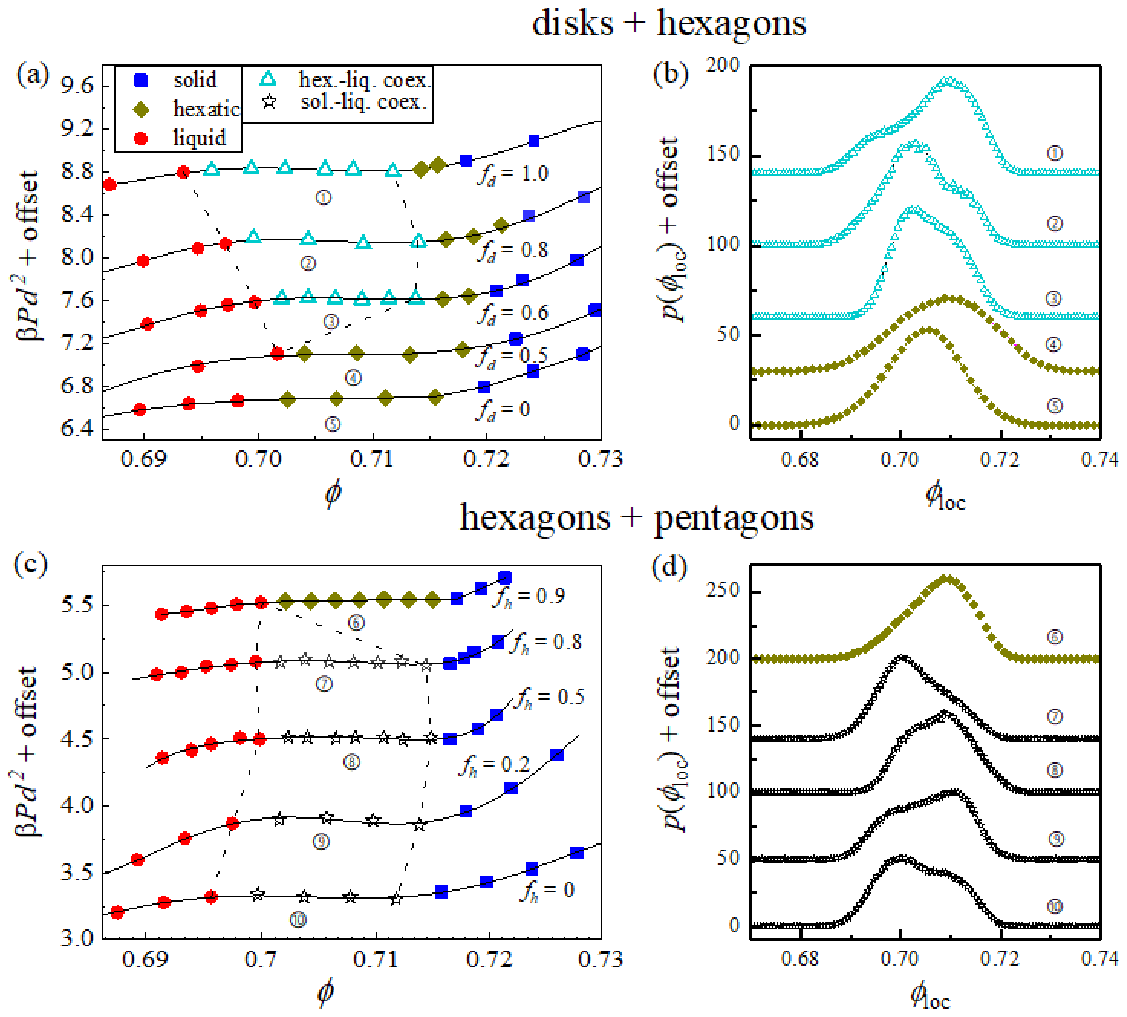}
 \caption{
Melting of the disks+hexagons mixture ((a) and (b)) and of the hexagons+pentagons mixture ((c) and (d)).
(a) and (c) illustrate the equation of state, and (b) and (d) show the probability distribution of the local density for selected state points highlighted in (a) and (c).
\label{fig:hdhp}
}
\end{figure}

Figure~\ref{fig:hdhp}(a) illustrates the isothermal equation of state of disks+hexagons for different disk fractions, and Fig.~\ref{fig:hdhp}(b) the probability distribution of local density for selected state points.
We use different symbols to identify the pure phases determined by studying the correlation functions $c(r)$ and $g_6(r)$ as in the main text.
The melting scenario crossover from the KTHNY of pure hexagons to the mixed of pure disks as the fraction of disks increases.
Consistently, the distribution of the local density transforms from unimodal to bimodal (Fig.~\ref{fig:hdhp}(b)).
Figs.~\ref{fig:hdhp}(c) and \ref{fig:hdhp}(d) similarly show a crossover from discontinuous melting to KTHNY on increasing the fraction of hexagons in hexagons+pentagons mixtures.

\newpage
\section{Melting of ternary mixtures \label{sec:ternery}}
\begin{figure*}[!h]
 \centering
 \includegraphics[angle=0,width=0.88\textwidth]{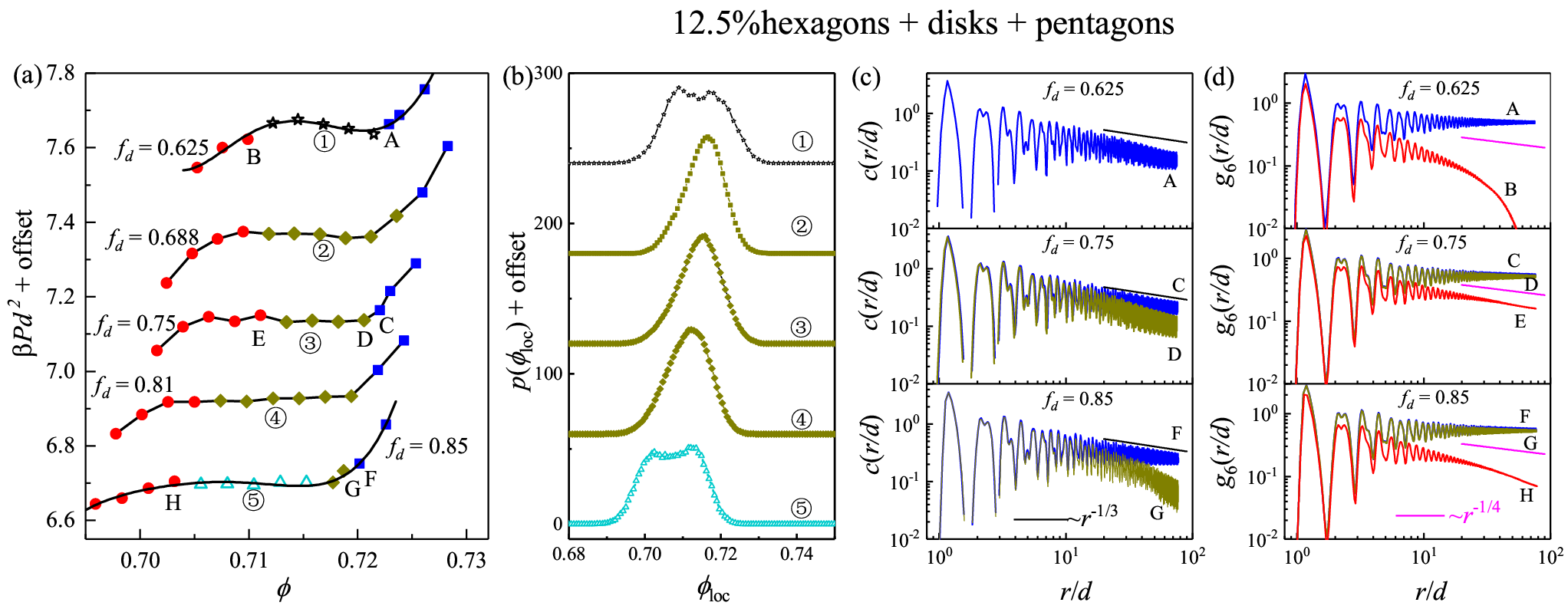}
 \caption{
Melting in ternary mixtures of pentagons, discs and hexagons. The fraction of hexagons is fixed at $0.125$.
(a) Isothermal equation of state. The disk fraction increases from top to bottom. 
Different symbols illustrate different phases, as detailed in the legend of Fig.~\ref{fig:hdhp}(a).
(b) probability distribution of the local density, and (c) the translational and (d) the bond-orientational correlation functions for selected state points marked in (a).
\label{fig:0.125hex}
}
\end{figure*}

\begin{figure*}[!h]
 \centering
 \includegraphics[angle=0,width=0.88\textwidth]{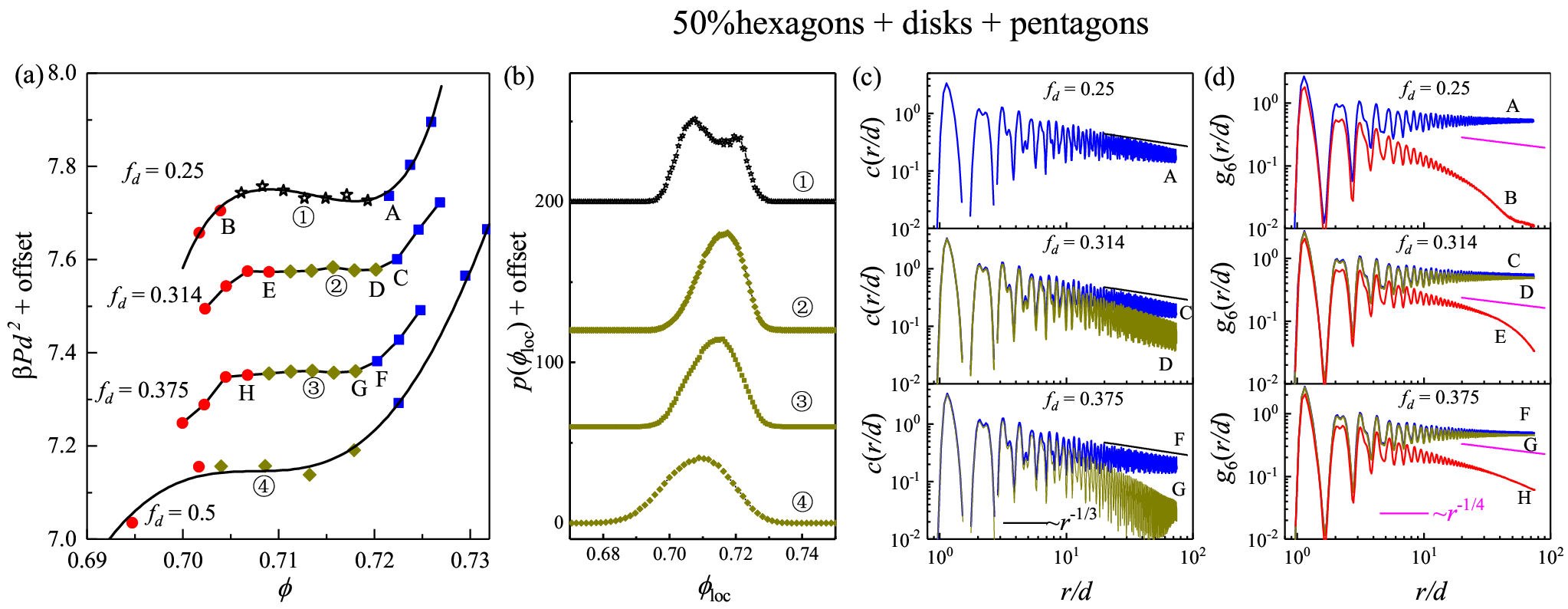}
 \caption{
The study of Fig.~\ref{fig:0.125hex} for $f_d = 0.125$ is repeated here for $f_d=0.5$.
\label{fig:0.5hex}
}
\end{figure*}

We illustrate in the following the studies we have conducted to determine the phase diagram of the ternary mixture of hexagons, pentagons and disks illustrated in Fig. 3(a) in the main text. 
We first fix $f_h=0.125$ and consider the melting scenario as the relative fraction of $f_p$ and $f_d$ changes.
Fig.~\ref{fig:0.125hex}(a) shows that the EOS has a Mayer-Wood loop at small and large $f_d$.
Correspondingly, the probability distribution of the local density is bimodal at small and large $f_d$ values, not at intermediate ones.
The translational and bond-orientational correlation functions, which we show in Figs.~\ref{fig:0.125hex}(c) and \ref{fig:0.125hex}(d), allow us to identify the pure phases.
The results demonstrate that, at fixed $f_h=0.125$, the melting scenario changes from first-order to KTHNY and finally to mixed on increasing $f_d$.
As a second example, we illustrate the determination of the phase behaviour at fixed $f_h=0.5$ in Fig.~\ref{fig:0.5hex}.
In this case, we observe a crossover from the first-order scenario to the KTHNY on increasing $f_d$.
The ternary phase diagram in Fig. 3(a) in the main text summarizes the results we obtained via similar studies performed at several $f_h$ values.

\newpage
\section{Defects \label{sec:pd}}

\begin{figure*}[b!]
 \centering
 \includegraphics[angle=0,width=0.8\textwidth]{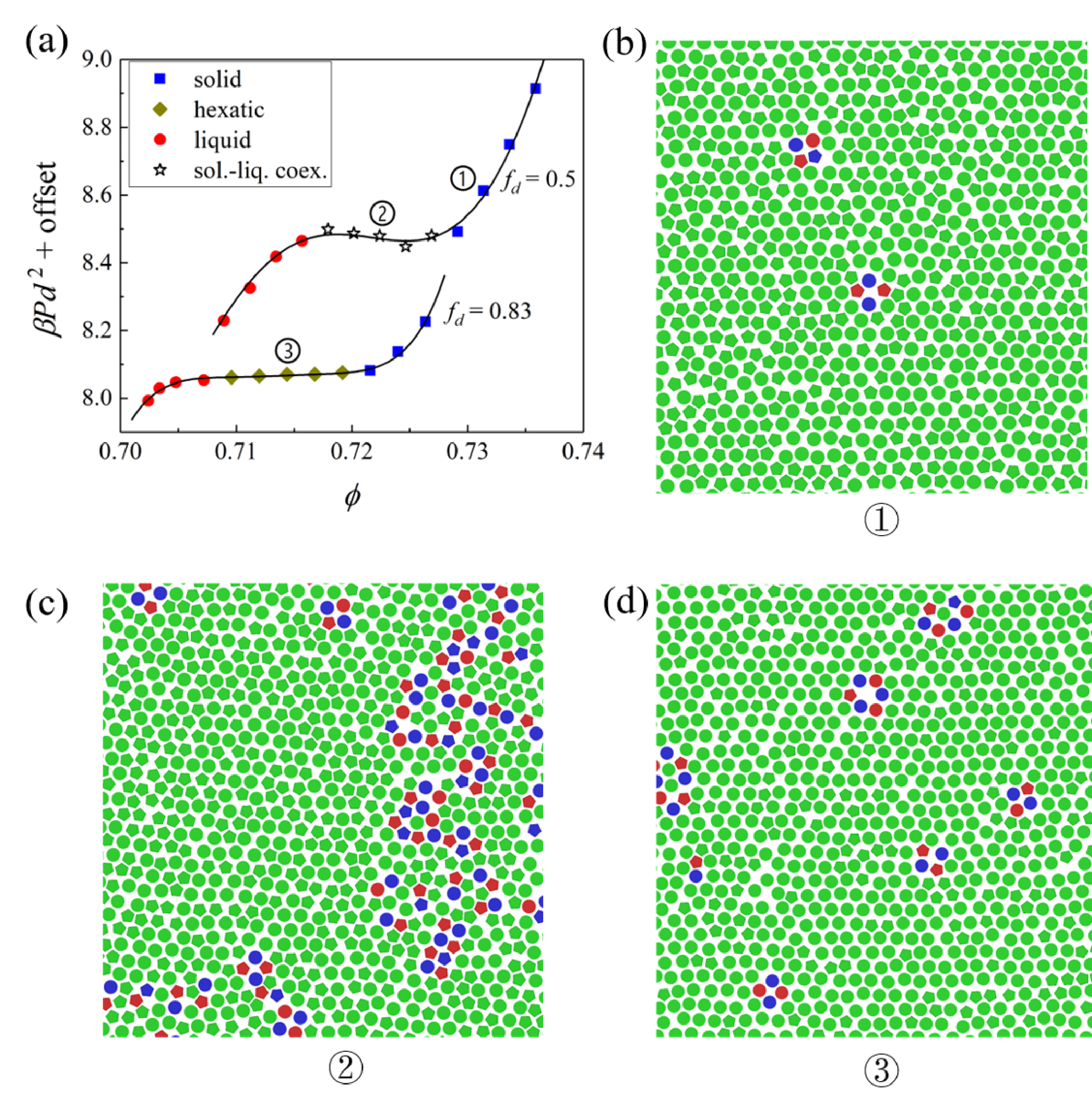}
\caption{(a) Equation of state of pentagons+disks mixture with $f_d=0.5$ and $0.83$. 
Panels (b),(c) and (d) are snapshots of a fraction of the system in the solid, solid-liquid coexistence and hexatic phases, as indicated in (a). 
Particles with 5, 6, and 7 neighbours are coloured in red, green, and blue, respectively.
\label{fig:snapshot}}
\end{figure*}

In two dimensions, melting and the KTHNY scenario~\cite{KT,HN,Y} have been related to the evolution of the density of topological defects, the unbinding of dislocation and disclination pairs driving the solid–hexatic and hexatic–liquid transitions. 
In Fig.~\ref{fig:snapshot}, we illustrate snapshots of pentagons+disks mixtures colour-coding the particles according to their number of Voronoi neighbours.
We illustrate results in the solid phase (panel (b), $f_d=0.5$), solid-liquid coexistence (panel (c), $f_d=0.5$), and in the hexatic phase (panel (d), $f_d=0.83$). The snapshots focus on a small region of our systems.
In the solid phase, we observe dislocation pairs. In the solid-liquid coexistence phase, we observe a region with a few defects and a region with many of them. Finally, the hexatic phases possess many isolated dislocations.

\begin{figure*}[h]
 \centering
 \includegraphics[angle=0,width=0.85\textwidth]{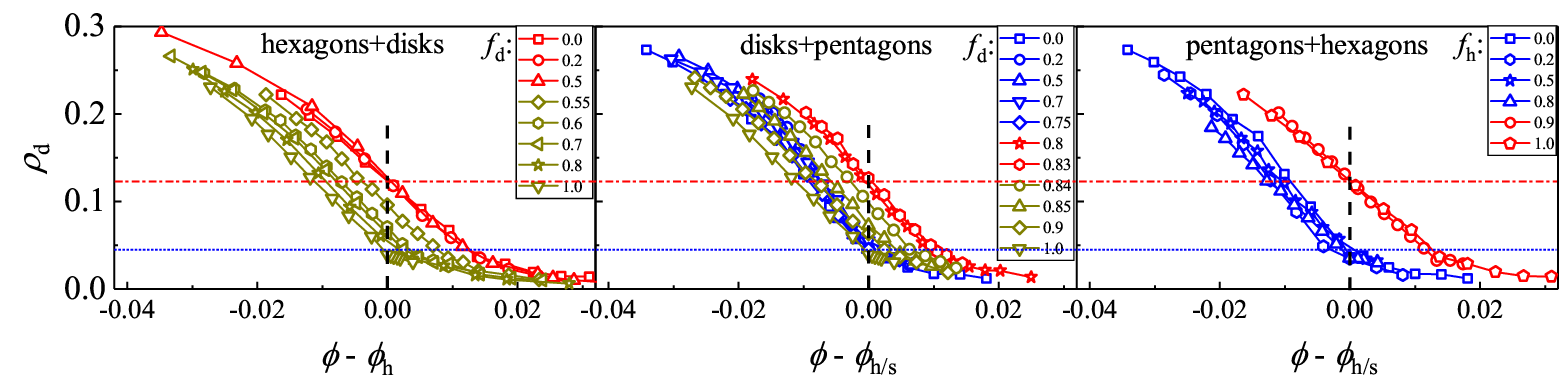}
\caption{Fraction of defects $\rho_d$ as a function of $\phi-\phi_h$, for system melting via the KTHNY (red) and mixed (yellow) scenario, and as a function of $\phi-\phi_s$ for systems melting via the first-order (blue) scenario.
Here, $\phi_h$ is the largest area fraction of the pure hexatic phase, and $\phi_s$ is the area fraction of pure solid at the end of liquid-solid coexistence for the first-order scenario.
The vertical dashed lines mark $\phi_h$ or $\phi_s$. 
The horizontal red dash-dotted line marks $\rho_d=0.123$, while the blue dotted line represents $\rho_d=0.046$.
\label{fig:defect}}
\end{figure*}

A recent work by Guo et al.~\cite{Guo2021} considered the crossover from discontinuous to mixed melting in bidisperse systems of disks differing in size or stiffness.
They found that when the system follows the KTHNY scenario, at the hexatic/liquid boundary the density of defective particles not having 6 neighbours approaches a universal value, $\rho_{d,{\rm h}}\simeq0.12$. 
We checked if this result also occurs in our system.
Figure~\ref{fig:defect} illustrates the area fraction dependence of $\rho_d$ in different mixtures. 
When the system follows the KTHNY melting scenario (red symbols), we find $\rho_{d,{\rm h}}=0.123\pm0.006$ (red dash-dotted line) at the hexatic/liquid boundary. 
As a comparison, we find $\rho_{d,{\rm h}}$ at the largest pure hexatic area fraction $\phi_h$ is not universal but depends on the relative fraction of two components if the system follows the mixed scenario (yellow symbols). 
These results are consistent with those previously reported~\cite{Guo2021}.

Figure~\ref{fig:defect} reveals that a related criterion characterizes discontinuous melting.
In the pure solid phase at the boundary between solid/liquid coexistence the density of defects acquires a universal value $\rho_{d,{\rm s}}=0.046\pm0.005$ (horizontal blue dotted line). 

\begin{figure*}[h]
 \centering
 \includegraphics[angle=0,width=0.75\textwidth]{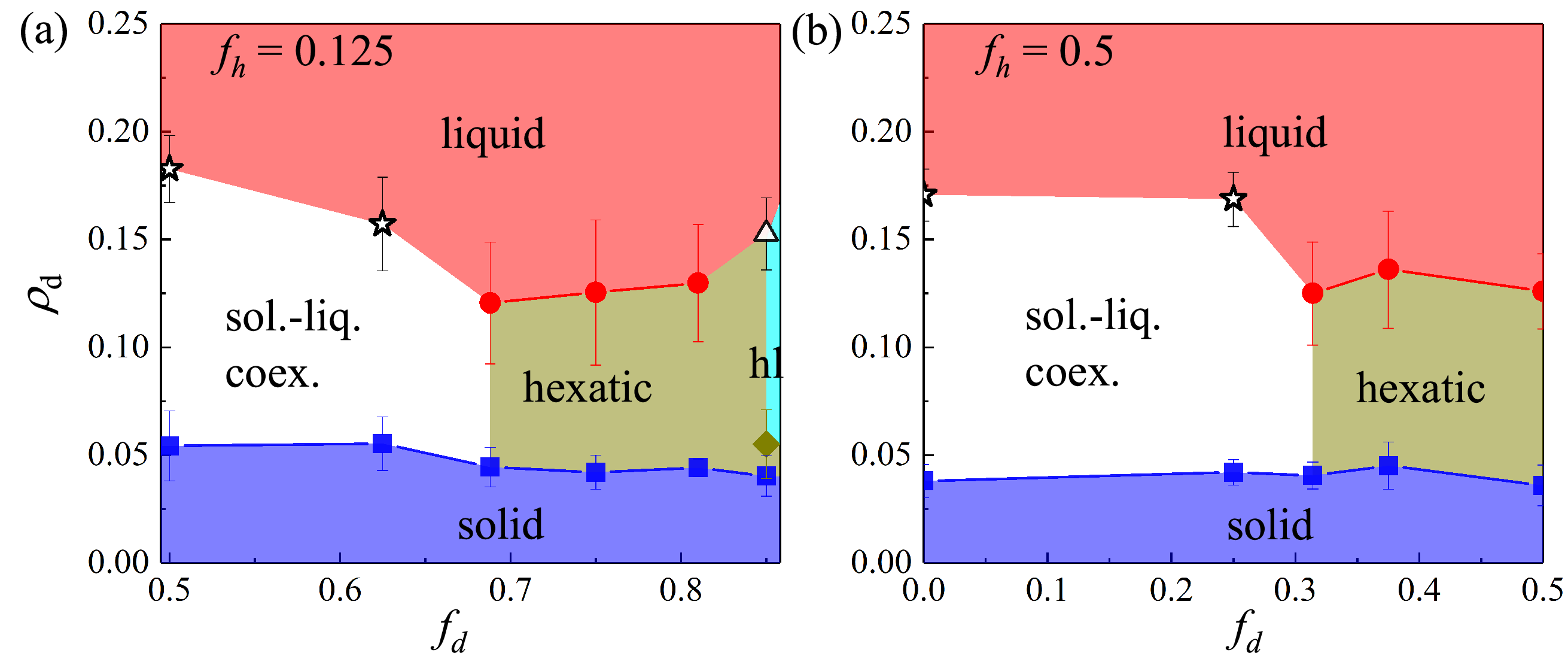}
\caption{The same plot as Fig. 3(b) in the main text, for the ternary system with (a) $f_h=0.125$ and (b) $f_h=0.5$.
\label{fig:defect_ternary}}
\end{figure*}

In the main text, we illustrated the composition dependence of the density of defects $\rho_d$ at the phase boundaries of binary mixtures, Fig.~2(b), and of the ternary mixture with $f_h=0.25$, in Fig.~3(b).
We investigate two additional $f_h$ values in the ternary system in Fig.~\ref{fig:defect_ternary}. 
In all considered cases, the solid phase vanishes when $\rho_{d,s}\simeq 0.046$, and the hexatic one when $\rho_{d,h}\simeq 0.123$.

\input{SM.bbl}

%% file: try.bbl
%

%% file: SM.bbl
%

%% file: try.bbl
\begin{thebibliography}{41}%
\makeatletter
\providecommand \@ifxundefined [1]{%
 \@ifx{#1\undefined}
}%
\providecommand \@ifnum [1]{%
 \ifnum #1\expandafter \@firstoftwo
 \else \expandafter \@secondoftwo
 \fi
}%
\providecommand \@ifx [1]{%
 \ifx #1\expandafter \@firstoftwo
 \else \expandafter \@secondoftwo
 \fi
}%
\providecommand \natexlab [1]{#1}%
\providecommand \enquote  [1]{``#1''}%
\providecommand \bibnamefont  [1]{#1}%
\providecommand \bibfnamefont [1]{#1}%
\providecommand \citenamefont [1]{#1}%
\providecommand \href@noop [0]{\@secondoftwo}%
\providecommand \href [0]{\begingroup \@sanitize@url \@href}%
\providecommand \@href[1]{\@@startlink{#1}\@@href}%
\providecommand \@@href[1]{\endgroup#1\@@endlink}%
\providecommand \@sanitize@url [0]{\catcode `\\12\catcode `\$12\catcode
  `\&12\catcode `\#12\catcode `\^12\catcode `\_12\catcode `\%12\relax}%
\providecommand \@@startlink[1]{}%
\providecommand \@@endlink[0]{}%
\providecommand \url  [0]{\begingroup\@sanitize@url \@url }%
\providecommand \@url [1]{\endgroup\@href {#1}{\urlprefix }}%
\providecommand \urlprefix  [0]{URL }%
\providecommand \Eprint [0]{\href }%
\providecommand \doibase [0]{http://dx.doi.org/}%
\providecommand \selectlanguage [0]{\@gobble}%
\providecommand \bibinfo  [0]{\@secondoftwo}%
\providecommand \bibfield  [0]{\@secondoftwo}%
\providecommand \translation [1]{[#1]}%
\providecommand \BibitemOpen [0]{}%
\providecommand \bibitemStop [0]{}%
\providecommand \bibitemNoStop [0]{.\EOS\space}%
\providecommand \EOS [0]{\spacefactor3000\relax}%
\providecommand \BibitemShut  [1]{\csname bibitem#1\endcsname}%
\let\auto@bib@innerbib\@empty
\bibitem [{\citenamefont {Kosterlitz}\ and\ \citenamefont
  {Thouless}(1973)}]{KT}%
  \BibitemOpen
  \bibfield  {author} {\bibinfo {author} {\bibfnamefont {J.~M.}\ \bibnamefont
  {Kosterlitz}}\ and\ \bibinfo {author} {\bibfnamefont {D.~J.}\ \bibnamefont
  {Thouless}},\ }\href {\doibase 10.1088/0022-3719/6/7/010} {\bibfield
  {journal} {\bibinfo  {journal} {J. Phys. C}\ }\textbf {\bibinfo {volume}
  {6}},\ \bibinfo {pages} {1181} (\bibinfo {year} {1973})}\BibitemShut
  {NoStop}%
\bibitem [{\citenamefont {Halperin}\ and\ \citenamefont {Nelson}(1978)}]{HN}%
  \BibitemOpen
  \bibfield  {author} {\bibinfo {author} {\bibfnamefont {B.~I.}\ \bibnamefont
  {Halperin}}\ and\ \bibinfo {author} {\bibfnamefont {D.~R.}\ \bibnamefont
  {Nelson}},\ }\href {\doibase 10.1103/PhysRevLett.41.121} {\bibfield
  {journal} {\bibinfo  {journal} {Phys. Rev. Lett.}\ }\textbf {\bibinfo
  {volume} {41}},\ \bibinfo {pages} {121} (\bibinfo {year} {1978})}\BibitemShut
  {NoStop}%
\bibitem [{\citenamefont {Young}(1979)}]{Y}%
  \BibitemOpen
  \bibfield  {author} {\bibinfo {author} {\bibfnamefont {A.~P.}\ \bibnamefont
  {Young}},\ }\href {\doibase 10.1103/PhysRevB.19.1855} {\bibfield  {journal}
  {\bibinfo  {journal} {Phys. Rev. B}\ }\textbf {\bibinfo {volume} {19}},\
  \bibinfo {pages} {1855} (\bibinfo {year} {1979})}\BibitemShut {NoStop}%
\bibitem [{\citenamefont {Brinkman}\ \emph {et~al.}(1982)\citenamefont
  {Brinkman}, \citenamefont {Fisher},\ and\ \citenamefont
  {Moncton}}]{Brinkman1982}%
  \BibitemOpen
  \bibfield  {author} {\bibinfo {author} {\bibfnamefont {W.~F.}\ \bibnamefont
  {Brinkman}}, \bibinfo {author} {\bibfnamefont {D.~S.}\ \bibnamefont
  {Fisher}}, \ and\ \bibinfo {author} {\bibfnamefont {D.~E.}\ \bibnamefont
  {Moncton}},\ }\href {\doibase 10.1126/science.217.4561.69} {\bibfield
  {journal} {\bibinfo  {journal} {Science}\ }\textbf {\bibinfo {volume}
  {217}},\ \bibinfo {pages} {693} (\bibinfo {year} {1982})}\BibitemShut
  {NoStop}%
\bibitem [{\citenamefont {Heiney}\ \emph {et~al.}(1983)\citenamefont {Heiney},
  \citenamefont {Stephens}, \citenamefont {Birgeneau}, \citenamefont {Horn},\
  and\ \citenamefont {Moncton}}]{Heiney}%
  \BibitemOpen
  \bibfield  {author} {\bibinfo {author} {\bibfnamefont {P.~A.}\ \bibnamefont
  {Heiney}}, \bibinfo {author} {\bibfnamefont {P.~W.}\ \bibnamefont
  {Stephens}}, \bibinfo {author} {\bibfnamefont {R.~J.}\ \bibnamefont
  {Birgeneau}}, \bibinfo {author} {\bibfnamefont {P.~M.}\ \bibnamefont {Horn}},
  \ and\ \bibinfo {author} {\bibfnamefont {D.~E.}\ \bibnamefont {Moncton}},\
  }\href {\doibase 10.1103/PhysRevB.28.6416} {\bibfield  {journal} {\bibinfo
  {journal} {Phys. Rev. B}\ }\textbf {\bibinfo {volume} {28}},\ \bibinfo
  {pages} {6416} (\bibinfo {year} {1983})}\BibitemShut {NoStop}%
\bibitem [{\citenamefont {Strandburg}(1988)}]{Strandburg}%
  \BibitemOpen
  \bibfield  {author} {\bibinfo {author} {\bibfnamefont {K.~J.}\ \bibnamefont
  {Strandburg}},\ }\href {\doibase 10.1103/RevModPhys.60.161} {\bibfield
  {journal} {\bibinfo  {journal} {Rev. Mod. Phys.}\ }\textbf {\bibinfo {volume}
  {60}},\ \bibinfo {pages} {161} (\bibinfo {year} {1988})}\BibitemShut
  {NoStop}%
\bibitem [{\citenamefont {Barker}\ \emph {et~al.}(1981)\citenamefont {Barker},
  \citenamefont {Henderson},\ and\ \citenamefont {Abraham}}]{LJpd1981}%
  \BibitemOpen
  \bibfield  {author} {\bibinfo {author} {\bibfnamefont {J.~A.}\ \bibnamefont
  {Barker}}, \bibinfo {author} {\bibfnamefont {D.}~\bibnamefont {Henderson}}, \
  and\ \bibinfo {author} {\bibfnamefont {F.~F.}\ \bibnamefont {Abraham}},\
  }\href {\doibase https://doi.org/10.1016/0378-4371(81)90222-3} {\bibfield
  {journal} {\bibinfo  {journal} {Physica A}\ }\textbf {\bibinfo {volume}
  {106}},\ \bibinfo {pages} {226} (\bibinfo {year} {1981})}\BibitemShut
  {NoStop}%
\bibitem [{\citenamefont {Bernard}\ and\ \citenamefont
  {Krauth}(2011)}]{Krauth2011}%
  \BibitemOpen
  \bibfield  {author} {\bibinfo {author} {\bibfnamefont {E.~P.}\ \bibnamefont
  {Bernard}}\ and\ \bibinfo {author} {\bibfnamefont {W.}~\bibnamefont
  {Krauth}},\ }\href {\doibase 10.1103/PhysRevLett.107.155704} {\bibfield
  {journal} {\bibinfo  {journal} {Phys. Rev. Lett.}\ }\textbf {\bibinfo
  {volume} {107}},\ \bibinfo {pages} {155704} (\bibinfo {year}
  {2011})}\BibitemShut {NoStop}%
\bibitem [{\citenamefont {Thorneywork}\ \emph {et~al.}(2017)\citenamefont
  {Thorneywork}, \citenamefont {Abbott}, \citenamefont {Aarts},\ and\
  \citenamefont {Dullens}}]{Experiment_harddisc}%
  \BibitemOpen
  \bibfield  {author} {\bibinfo {author} {\bibfnamefont {A.~L.}\ \bibnamefont
  {Thorneywork}}, \bibinfo {author} {\bibfnamefont {J.~L.}\ \bibnamefont
  {Abbott}}, \bibinfo {author} {\bibfnamefont {D.~G. A.~L.}\ \bibnamefont
  {Aarts}}, \ and\ \bibinfo {author} {\bibfnamefont {R.~P.~A.}\ \bibnamefont
  {Dullens}},\ }\href {\doibase 10.1103/PhysRevLett.118.158001} {\bibfield
  {journal} {\bibinfo  {journal} {Phys. Rev. Lett.}\ }\textbf {\bibinfo
  {volume} {118}},\ \bibinfo {pages} {158001} (\bibinfo {year}
  {2017})}\BibitemShut {NoStop}%
\bibitem [{\citenamefont {Li}\ and\ \citenamefont
  {Ciamarra}(2020)}]{Massimo_2020PRLmelting}%
  \BibitemOpen
  \bibfield  {author} {\bibinfo {author} {\bibfnamefont {Y.-W.}\ \bibnamefont
  {Li}}\ and\ \bibinfo {author} {\bibfnamefont {M.~P.}\ \bibnamefont
  {Ciamarra}},\ }\href {\doibase 10.1103/PhysRevLett.124.218002} {\bibfield
  {journal} {\bibinfo  {journal} {Phys. Rev. Lett.}\ }\textbf {\bibinfo
  {volume} {124}},\ \bibinfo {pages} {218002} (\bibinfo {year}
  {2020})}\BibitemShut {NoStop}%
\bibitem [{\citenamefont {Komatsu}\ and\ \citenamefont
  {Tanaka}(2015)}]{Tanaka}%
  \BibitemOpen
  \bibfield  {author} {\bibinfo {author} {\bibfnamefont {Y.}~\bibnamefont
  {Komatsu}}\ and\ \bibinfo {author} {\bibfnamefont {H.}~\bibnamefont
  {Tanaka}},\ }\href {\doibase 10.1103/PhysRevX.5.031025} {\bibfield  {journal}
  {\bibinfo  {journal} {Phys. Rev. X}\ }\textbf {\bibinfo {volume} {5}},\
  \bibinfo {pages} {031025} (\bibinfo {year} {2015})}\BibitemShut {NoStop}%
\bibitem [{\citenamefont {Li}\ and\ \citenamefont
  {Pica~Ciamarra}(2018)}]{Cell}%
  \BibitemOpen
  \bibfield  {author} {\bibinfo {author} {\bibfnamefont {Y.-W.}\ \bibnamefont
  {Li}}\ and\ \bibinfo {author} {\bibfnamefont {M.}~\bibnamefont
  {Pica~Ciamarra}},\ }\href {\doibase 10.1103/PhysRevMaterials.2.045602}
  {\bibfield  {journal} {\bibinfo  {journal} {Phys. Rev. Mater.}\ }\textbf
  {\bibinfo {volume} {2}},\ \bibinfo {pages} {045602} (\bibinfo {year}
  {2018})}\BibitemShut {NoStop}%
\bibitem [{\citenamefont {Deutschl\"{a}nder}\ \emph {et~al.}(2013)\citenamefont
  {Deutschl\"{a}nder}, \citenamefont {Horn}, \citenamefont {L\"{o}wen},
  \citenamefont {Maret},\ and\ \citenamefont {Keim}}]{pinned_particle}%
  \BibitemOpen
  \bibfield  {author} {\bibinfo {author} {\bibfnamefont {S.}~\bibnamefont
  {Deutschl\"{a}nder}}, \bibinfo {author} {\bibfnamefont {T.}~\bibnamefont
  {Horn}}, \bibinfo {author} {\bibfnamefont {H.}~\bibnamefont {L\"{o}wen}},
  \bibinfo {author} {\bibfnamefont {G.}~\bibnamefont {Maret}}, \ and\ \bibinfo
  {author} {\bibfnamefont {P.}~\bibnamefont {Keim}},\ }\href {\doibase
  10.1103/PhysRevLett.111.098301} {\bibfield  {journal} {\bibinfo  {journal}
  {Phys. Rev. Lett.}\ }\textbf {\bibinfo {volume} {111}},\ \bibinfo {pages}
  {098301} (\bibinfo {year} {2013})}\BibitemShut {NoStop}%
\bibitem [{\citenamefont {Downs}\ \emph {et~al.}(2021)\citenamefont {Downs},
  \citenamefont {Smith}, \citenamefont {Mandadapu}, \citenamefont {Garrahan},\
  and\ \citenamefont {Smith}}]{Smith2021}%
  \BibitemOpen
  \bibfield  {author} {\bibinfo {author} {\bibfnamefont {J.}~\bibnamefont
  {Downs}}, \bibinfo {author} {\bibfnamefont {N.}~\bibnamefont {Smith}},
  \bibinfo {author} {\bibfnamefont {K.}~\bibnamefont {Mandadapu}}, \bibinfo
  {author} {\bibfnamefont {J.~P.}\ \bibnamefont {Garrahan}}, \ and\ \bibinfo
  {author} {\bibfnamefont {M.~I.}\ \bibnamefont {Smith}},\ }\href {\doibase
  10.1103/PhysRevLett.127.268002} {\bibfield  {journal} {\bibinfo  {journal}
  {Phys. Rev. Lett.}\ }\textbf {\bibinfo {volume} {127}},\ \bibinfo {pages}
  {268002} (\bibinfo {year} {2021})}\BibitemShut {NoStop}%
\bibitem [{\citenamefont {Deutschländer}\ \emph {et~al.}(2014)\citenamefont
  {Deutschländer}, \citenamefont {Puertas}, \citenamefont {Maret},\ and\
  \citenamefont {Keim}}]{Keim}%
  \BibitemOpen
  \bibfield  {author} {\bibinfo {author} {\bibfnamefont {S.}~\bibnamefont
  {Deutschländer}}, \bibinfo {author} {\bibfnamefont {A.~M.}\ \bibnamefont
  {Puertas}}, \bibinfo {author} {\bibfnamefont {G.}~\bibnamefont {Maret}}, \
  and\ \bibinfo {author} {\bibfnamefont {P.}~\bibnamefont {Keim}},\ }\href
  {\doibase 10.1103/PhysRevLett.113.127801} {\bibfield  {journal} {\bibinfo
  {journal} {Phys. Rev. Lett.}\ }\textbf {\bibinfo {volume} {113}},\ \bibinfo
  {pages} {127801} (\bibinfo {year} {2014})}\BibitemShut {NoStop}%
\bibitem [{\citenamefont {Abutbul}\ and\ \citenamefont
  {Podolsky}(2022)}]{Daniel2022}%
  \BibitemOpen
  \bibfield  {author} {\bibinfo {author} {\bibfnamefont {D.}~\bibnamefont
  {Abutbul}}\ and\ \bibinfo {author} {\bibfnamefont {D.}~\bibnamefont
  {Podolsky}},\ }\href {\doibase 10.1103/PhysRevLett.128.255501} {\bibfield
  {journal} {\bibinfo  {journal} {Phys. Rev. Lett.}\ }\textbf {\bibinfo
  {volume} {128}},\ \bibinfo {pages} {255501} (\bibinfo {year}
  {2022})}\BibitemShut {NoStop}%
\bibitem [{\citenamefont {Kapfer}\ and\ \citenamefont
  {Krauth}(2015)}]{Krauth2015}%
  \BibitemOpen
  \bibfield  {author} {\bibinfo {author} {\bibfnamefont {S.~C.}\ \bibnamefont
  {Kapfer}}\ and\ \bibinfo {author} {\bibfnamefont {W.}~\bibnamefont
  {Krauth}},\ }\href {\doibase 10.1103/PhysRevLett.114.035702} {\bibfield
  {journal} {\bibinfo  {journal} {Phys. Rev. Lett.}\ }\textbf {\bibinfo
  {volume} {114}},\ \bibinfo {pages} {035702} (\bibinfo {year}
  {2015})}\BibitemShut {NoStop}%
\bibitem [{\citenamefont {Anderson}\ \emph {et~al.}(2017)\citenamefont
  {Anderson}, \citenamefont {Antonaglia}, \citenamefont {Millan}, \citenamefont
  {Engel},\ and\ \citenamefont {Glotzer}}]{Glotzer}%
  \BibitemOpen
  \bibfield  {author} {\bibinfo {author} {\bibfnamefont {J.~A.}\ \bibnamefont
  {Anderson}}, \bibinfo {author} {\bibfnamefont {J.}~\bibnamefont
  {Antonaglia}}, \bibinfo {author} {\bibfnamefont {J.~A.}\ \bibnamefont
  {Millan}}, \bibinfo {author} {\bibfnamefont {M.}~\bibnamefont {Engel}}, \
  and\ \bibinfo {author} {\bibfnamefont {S.~C.}\ \bibnamefont {Glotzer}},\
  }\href {\doibase 10.1103/PhysRevX.7.021001} {\bibfield  {journal} {\bibinfo
  {journal} {Phys. Rev. X}\ }\textbf {\bibinfo {volume} {7}},\ \bibinfo {pages}
  {021001} (\bibinfo {year} {2017})}\BibitemShut {NoStop}%
\bibitem [{\citenamefont {Prestipino}\ \emph {et~al.}(2011)\citenamefont
  {Prestipino}, \citenamefont {Saija},\ and\ \citenamefont
  {Giaquinta}}]{Santi}%
  \BibitemOpen
  \bibfield  {author} {\bibinfo {author} {\bibfnamefont {S.}~\bibnamefont
  {Prestipino}}, \bibinfo {author} {\bibfnamefont {F.}~\bibnamefont {Saija}}, \
  and\ \bibinfo {author} {\bibfnamefont {P.~V.}\ \bibnamefont {Giaquinta}},\
  }\href {\doibase 10.1103/PhysRevLett.106.235701} {\bibfield  {journal}
  {\bibinfo  {journal} {Phys. Rev. Lett.}\ }\textbf {\bibinfo {volume} {106}},\
  \bibinfo {pages} {235701} (\bibinfo {year} {2011})}\BibitemShut {NoStop}%
\bibitem [{\citenamefont {Zu}\ \emph {et~al.}(2016)\citenamefont {Zu},
  \citenamefont {Liu}, \citenamefont {Tong},\ and\ \citenamefont
  {Xu}}]{ningxu}%
  \BibitemOpen
  \bibfield  {author} {\bibinfo {author} {\bibfnamefont {M.}~\bibnamefont
  {Zu}}, \bibinfo {author} {\bibfnamefont {J.}~\bibnamefont {Liu}}, \bibinfo
  {author} {\bibfnamefont {H.}~\bibnamefont {Tong}}, \ and\ \bibinfo {author}
  {\bibfnamefont {N.}~\bibnamefont {Xu}},\ }\href {\doibase
  10.1103/PhysRevLett.117.085702} {\bibfield  {journal} {\bibinfo  {journal}
  {Phys. Rev. Lett.}\ }\textbf {\bibinfo {volume} {117}},\ \bibinfo {pages}
  {85702} (\bibinfo {year} {2016})}\BibitemShut {NoStop}%
\bibitem [{\citenamefont {Russo}\ and\ \citenamefont
  {Wilding}(2017)}]{John_Russo}%
  \BibitemOpen
  \bibfield  {author} {\bibinfo {author} {\bibfnamefont {J.}~\bibnamefont
  {Russo}}\ and\ \bibinfo {author} {\bibfnamefont {N.~B.}\ \bibnamefont
  {Wilding}},\ }\href {\doibase 10.1103/PhysRevLett.119.115702} {\bibfield
  {journal} {\bibinfo  {journal} {Phys. Rev. Lett.}\ }\textbf {\bibinfo
  {volume} {119}},\ \bibinfo {pages} {115702} (\bibinfo {year}
  {2017})}\BibitemShut {NoStop}%
\bibitem [{\citenamefont {Guo}\ \emph {et~al.}(2021)\citenamefont {Guo},
  \citenamefont {Nie},\ and\ \citenamefont {Xu}}]{Guo2021}%
  \BibitemOpen
  \bibfield  {author} {\bibinfo {author} {\bibfnamefont {J.}~\bibnamefont
  {Guo}}, \bibinfo {author} {\bibfnamefont {Y.}~\bibnamefont {Nie}}, \ and\
  \bibinfo {author} {\bibfnamefont {N.}~\bibnamefont {Xu}},\ }\href {\doibase
  10.1039/D0SM02199G} {\bibfield  {journal} {\bibinfo  {journal} {Soft Matter}\
  }\textbf {\bibinfo {volume} {17}},\ \bibinfo {pages} {3397} (\bibinfo {year}
  {2021})}\BibitemShut {NoStop}%
\bibitem [{\citenamefont {Sampedro~Ruiz}\ \emph {et~al.}(2019)\citenamefont
  {Sampedro~Ruiz}, \citenamefont {Lei},\ and\ \citenamefont {Ni}}]{Ran2dmelt}%
  \BibitemOpen
  \bibfield  {author} {\bibinfo {author} {\bibfnamefont {P.}~\bibnamefont
  {Sampedro~Ruiz}}, \bibinfo {author} {\bibfnamefont {Q.-l.}\ \bibnamefont
  {Lei}}, \ and\ \bibinfo {author} {\bibfnamefont {R.}~\bibnamefont {Ni}},\
  }\href {\doibase 10.1038/s42005-019-0172-2} {\bibfield  {journal} {\bibinfo
  {journal} {Commun. Phys.}\ }\textbf {\bibinfo {volume} {2}},\ \bibinfo
  {pages} {70} (\bibinfo {year} {2019})}\BibitemShut {NoStop}%
\bibitem [{\citenamefont {Digregorio}\ \emph {et~al.}(2022)\citenamefont
  {Digregorio}, \citenamefont {Levis}, \citenamefont {Cugliandolo},
  \citenamefont {Gonnella},\ and\ \citenamefont {Pagonabarraga}}]{defects_sm}%
  \BibitemOpen
  \bibfield  {author} {\bibinfo {author} {\bibfnamefont {P.}~\bibnamefont
  {Digregorio}}, \bibinfo {author} {\bibfnamefont {D.}~\bibnamefont {Levis}},
  \bibinfo {author} {\bibfnamefont {L.~F.}\ \bibnamefont {Cugliandolo}},
  \bibinfo {author} {\bibfnamefont {G.}~\bibnamefont {Gonnella}}, \ and\
  \bibinfo {author} {\bibfnamefont {I.}~\bibnamefont {Pagonabarraga}},\ }\href
  {\doibase 10.1039/D1SM01411K} {\bibfield  {journal} {\bibinfo  {journal}
  {Soft Matter}\ }\textbf {\bibinfo {volume} {18}},\ \bibinfo {pages} {566}
  (\bibinfo {year} {2022})}\BibitemShut {NoStop}%
\bibitem [{\citenamefont {Zahn}\ and\ \citenamefont {Maret}(2000)}]{Zahn2000}%
  \BibitemOpen
  \bibfield  {author} {\bibinfo {author} {\bibfnamefont {K.}~\bibnamefont
  {Zahn}}\ and\ \bibinfo {author} {\bibfnamefont {G.}~\bibnamefont {Maret}},\
  }\href {\doibase 10.1103/PhysRevLett.85.3656} {\bibfield  {journal} {\bibinfo
   {journal} {Phys. Rev. Lett.}\ }\textbf {\bibinfo {volume} {85}},\ \bibinfo
  {pages} {3656} (\bibinfo {year} {2000})}\BibitemShut {NoStop}%
\bibitem [{\citenamefont {Khrapak}(2020)}]{Khrapak2020}%
  \BibitemOpen
  \bibfield  {author} {\bibinfo {author} {\bibfnamefont {S.~A.}\ \bibnamefont
  {Khrapak}},\ }\href {\doibase 10.1103/PhysRevResearch.2.012040} {\bibfield
  {journal} {\bibinfo  {journal} {Phys. Rev. Res.}\ }\textbf {\bibinfo {volume}
  {2}},\ \bibinfo {pages} {012040} (\bibinfo {year} {2020})}\BibitemShut
  {NoStop}%
\bibitem [{\citenamefont {Dillmann}\ \emph {et~al.}(2012)\citenamefont
  {Dillmann}, \citenamefont {Maret},\ and\ \citenamefont
  {Keim}}]{Dillmann2012}%
  \BibitemOpen
  \bibfield  {author} {\bibinfo {author} {\bibfnamefont {P.}~\bibnamefont
  {Dillmann}}, \bibinfo {author} {\bibfnamefont {G.}~\bibnamefont {Maret}}, \
  and\ \bibinfo {author} {\bibfnamefont {P.}~\bibnamefont {Keim}},\ }\href
  {\doibase 10.1088/0953-8984/24/46/464118} {\bibfield  {journal} {\bibinfo
  {journal} {J. Phys. Condens. Matter}\ }\textbf {\bibinfo {volume} {24}},\
  \bibinfo {pages} {464118} (\bibinfo {year} {2012})}\BibitemShut {NoStop}%
\bibitem [{\citenamefont {Zong}\ and\ \citenamefont {Zhao}(2022)}]{Zong2022}%
  \BibitemOpen
  \bibfield  {author} {\bibinfo {author} {\bibfnamefont {Y.}~\bibnamefont
  {Zong}}\ and\ \bibinfo {author} {\bibfnamefont {K.}~\bibnamefont {Zhao}},\
  }\href {\doibase 10.1016/J.COSSMS.2022.101022} {\bibfield  {journal}
  {\bibinfo  {journal} {Curr. Opin. Solid State Mater. Sci.}\ }\textbf
  {\bibinfo {volume} {26}},\ \bibinfo {pages} {101022} (\bibinfo {year}
  {2022})}\BibitemShut {NoStop}%
\bibitem [{SM()}]{SM}%
  \BibitemOpen
  \href@noop {} {}\bibinfo {note} {See Supplemental Material at http://... for
  additional information about numerical details, thermal equilibration, finite
  size effects, temperature dependence of the melting, the melting of
  bidisperse and ternary systems, and defects.}\BibitemShut {Stop}%
\bibitem [{\citenamefont {Weeks}\ \emph {et~al.}(1971)\citenamefont {Weeks},
  \citenamefont {Chandler},\ and\ \citenamefont {Andersen}}]{wca}%
  \BibitemOpen
  \bibfield  {author} {\bibinfo {author} {\bibfnamefont {J.~D.}\ \bibnamefont
  {Weeks}}, \bibinfo {author} {\bibfnamefont {D.}~\bibnamefont {Chandler}}, \
  and\ \bibinfo {author} {\bibfnamefont {H.~C.}\ \bibnamefont {Andersen}},\
  }\href {\doibase 10.1063/1.1674820} {\bibfield  {journal} {\bibinfo
  {journal} {J. Chem. Phys.}\ }\textbf {\bibinfo {volume} {54}},\ \bibinfo
  {pages} {5237} (\bibinfo {year} {1971})}\BibitemShut {NoStop}%
\bibitem [{\citenamefont {Zhu}\ \emph {et~al.}(2013)\citenamefont {Zhu},
  \citenamefont {Liu}, \citenamefont {Li}, \citenamefont {Qian}, \citenamefont
  {Milano},\ and\ \citenamefont {Lu}}]{Galamost}%
  \BibitemOpen
  \bibfield  {author} {\bibinfo {author} {\bibfnamefont {Y.}~\bibnamefont
  {Zhu}}, \bibinfo {author} {\bibfnamefont {H.}~\bibnamefont {Liu}}, \bibinfo
  {author} {\bibfnamefont {Z.}~\bibnamefont {Li}}, \bibinfo {author}
  {\bibfnamefont {H.}~\bibnamefont {Qian}}, \bibinfo {author} {\bibfnamefont
  {G.}~\bibnamefont {Milano}}, \ and\ \bibinfo {author} {\bibfnamefont
  {Z.}~\bibnamefont {Lu}},\ }\href {\doibase https://doi.org/10.1002/jcc.23365}
  {\bibfield  {journal} {\bibinfo  {journal} {J. Comput. Chem.}\ }\textbf
  {\bibinfo {volume} {34}},\ \bibinfo {pages} {2197} (\bibinfo {year}
  {2013})}\BibitemShut {NoStop}%
\bibitem [{\citenamefont {Likos}\ and\ \citenamefont {Henley}(1993)}]{Likos93}%
  \BibitemOpen
  \bibfield  {author} {\bibinfo {author} {\bibfnamefont {C.~N.}\ \bibnamefont
  {Likos}}\ and\ \bibinfo {author} {\bibfnamefont {C.~L.}\ \bibnamefont
  {Henley}},\ }\href {\doibase 10.1080/13642819308215284} {\bibfield  {journal}
  {\bibinfo  {journal} {Philos. mag., B}\ }\textbf {\bibinfo {volume} {68}},\
  \bibinfo {pages} {85} (\bibinfo {year} {1993})}\BibitemShut {NoStop}%
\bibitem [{\citenamefont {Uche}\ \emph {et~al.}(2004)\citenamefont {Uche},
  \citenamefont {Stillinger},\ and\ \citenamefont {Torquato}}]{UCHE2004428}%
  \BibitemOpen
  \bibfield  {author} {\bibinfo {author} {\bibfnamefont {O.}~\bibnamefont
  {Uche}}, \bibinfo {author} {\bibfnamefont {F.}~\bibnamefont {Stillinger}}, \
  and\ \bibinfo {author} {\bibfnamefont {S.}~\bibnamefont {Torquato}},\ }\href
  {\doibase https://doi.org/10.1016/j.physa.2004.05.082} {\bibfield  {journal}
  {\bibinfo  {journal} {Phys. A: Stat. Mech.}\ }\textbf {\bibinfo {volume}
  {342}},\ \bibinfo {pages} {428} (\bibinfo {year} {2004})}\BibitemShut
  {NoStop}%
\bibitem [{\citenamefont {Perera-Burgos}\ \emph {et~al.}(2016)\citenamefont
  {Perera-Burgos}, \citenamefont {M{\'e}ndez-Alcaraz}, \citenamefont
  {P{\'e}rez-{\'A}ngel},\ and\ \citenamefont
  {Casta{\~n}eda-Priego}}]{Castaeda2016}%
  \BibitemOpen
  \bibfield  {author} {\bibinfo {author} {\bibfnamefont {J.~A.}\ \bibnamefont
  {Perera-Burgos}}, \bibinfo {author} {\bibfnamefont {J.~M.}\ \bibnamefont
  {M{\'e}ndez-Alcaraz}}, \bibinfo {author} {\bibfnamefont {G.}~\bibnamefont
  {P{\'e}rez-{\'A}ngel}}, \ and\ \bibinfo {author} {\bibfnamefont
  {R.}~\bibnamefont {Casta{\~n}eda-Priego}},\ }\href {\doibase
  10.1063/1.4962423} {\bibfield  {journal} {\bibinfo  {journal} {J. Chem.
  Phys.}\ }\textbf {\bibinfo {volume} {145}},\ \bibinfo {pages} {104905}
  (\bibinfo {year} {2016})}\BibitemShut {NoStop}%
\bibitem [{\citenamefont {Fayen}\ \emph {et~al.}(2020)\citenamefont {Fayen},
  \citenamefont {Jagannathan}, \citenamefont {Foffi},\ and\ \citenamefont
  {Smallenburg}}]{Smallenburg2020}%
  \BibitemOpen
  \bibfield  {author} {\bibinfo {author} {\bibfnamefont {E.}~\bibnamefont
  {Fayen}}, \bibinfo {author} {\bibfnamefont {A.}~\bibnamefont {Jagannathan}},
  \bibinfo {author} {\bibfnamefont {G.}~\bibnamefont {Foffi}}, \ and\ \bibinfo
  {author} {\bibfnamefont {F.}~\bibnamefont {Smallenburg}},\ }\href {\doibase
  10.1063/5.0008230} {\bibfield  {journal} {\bibinfo  {journal} {J. Chem.
  Phys.}\ }\textbf {\bibinfo {volume} {152}},\ \bibinfo {pages} {204901}
  (\bibinfo {year} {2020})}\BibitemShut {NoStop}%
\bibitem [{\citenamefont {Prajwal}\ and\ \citenamefont
  {Escobedo}(2021)}]{Escobedo2021}%
  \BibitemOpen
  \bibfield  {author} {\bibinfo {author} {\bibfnamefont {B.~P.}\ \bibnamefont
  {Prajwal}}\ and\ \bibinfo {author} {\bibfnamefont {F.~A.}\ \bibnamefont
  {Escobedo}},\ }\href {\doibase 10.1103/PhysRevMaterials.5.024003} {\bibfield
  {journal} {\bibinfo  {journal} {Phys. Rev. Mater.}\ }\textbf {\bibinfo
  {volume} {5}},\ \bibinfo {pages} {024003} (\bibinfo {year}
  {2021})}\BibitemShut {NoStop}%
\bibitem [{\citenamefont {Speedy}(1999)}]{Speedy1999}%
  \BibitemOpen
  \bibfield  {author} {\bibinfo {author} {\bibfnamefont {R.~J.}\ \bibnamefont
  {Speedy}},\ }\href {\doibase 10.1063/1.478337} {\bibfield  {journal}
  {\bibinfo  {journal} {J. Chem. Phys.}\ }\textbf {\bibinfo {volume} {110}},\
  \bibinfo {pages} {4559} (\bibinfo {year} {1999})}\BibitemShut {NoStop}%
\bibitem [{\citenamefont {Mayer}\ and\ \citenamefont
  {Wood}(1965)}]{Mayer_wood}%
  \BibitemOpen
  \bibfield  {author} {\bibinfo {author} {\bibfnamefont {J.~E.}\ \bibnamefont
  {Mayer}}\ and\ \bibinfo {author} {\bibfnamefont {W.~W.}\ \bibnamefont
  {Wood}},\ }\href {\doibase 10.1063/1.1695931} {\bibfield  {journal} {\bibinfo
   {journal} {J. Chem. Phys.}\ }\textbf {\bibinfo {volume} {42}},\ \bibinfo
  {pages} {4268} (\bibinfo {year} {1965})}\BibitemShut {NoStop}%
\bibitem [{\citenamefont {Li}\ and\ \citenamefont
  {Ciamarra}(2019)}]{Massimo_PRE2019}%
  \BibitemOpen
  \bibfield  {author} {\bibinfo {author} {\bibfnamefont {Y.-W.}\ \bibnamefont
  {Li}}\ and\ \bibinfo {author} {\bibfnamefont {M.~P.}\ \bibnamefont
  {Ciamarra}},\ }\href {\doibase 10.1103/PhysRevE.100.062606} {\bibfield
  {journal} {\bibinfo  {journal} {Phys. Rev. E}\ }\textbf {\bibinfo {volume}
  {100}},\ \bibinfo {pages} {062606} (\bibinfo {year} {2019})}\BibitemShut
  {NoStop}%
\bibitem [{\citenamefont {Mermin}\ and\ \citenamefont {Wagner}(1966)}]{Mermin}%
  \BibitemOpen
  \bibfield  {author} {\bibinfo {author} {\bibfnamefont {N.~D.}\ \bibnamefont
  {Mermin}}\ and\ \bibinfo {author} {\bibfnamefont {H.}~\bibnamefont
  {Wagner}},\ }\href {\doibase 10.1103/PhysRevLett.17.1133} {\bibfield
  {journal} {\bibinfo  {journal} {Phys. Rev. Lett.}\ }\textbf {\bibinfo
  {volume} {17}},\ \bibinfo {pages} {1133} (\bibinfo {year}
  {1966})}\BibitemShut {NoStop}%
\bibitem [{\citenamefont {Toyotama}\ \emph {et~al.}(2016)\citenamefont
  {Toyotama}, \citenamefont {Okuzono},\ and\ \citenamefont
  {Yamanaka}}]{Toyotama2016}%
  \BibitemOpen
  \bibfield  {author} {\bibinfo {author} {\bibfnamefont {A.}~\bibnamefont
  {Toyotama}}, \bibinfo {author} {\bibfnamefont {T.}~\bibnamefont {Okuzono}}, \
  and\ \bibinfo {author} {\bibfnamefont {J.}~\bibnamefont {Yamanaka}},\ }\href
  {\doibase 10.1038/srep23292} {\bibfield  {journal} {\bibinfo  {journal} {Sci.
  Rep.}\ }\textbf {\bibinfo {volume} {6}},\ \bibinfo {pages} {23292} (\bibinfo
  {year} {2016})}\BibitemShut {NoStop}%
\end{thebibliography}
